\documentclass[twocolumn,floatfix,prb,aps,showpacs]{revtex4-1}
\usepackage{graphicx,amsmath,amssymb}
\usepackage{multirow}
\usepackage{epstopdf}

\begin{document}

\title{Theory of inter-Landau level magnetoexcitons in bilayer graphene} 
\author{Judit S\'ari$^1$ and Csaba T\H oke$^{2}$}
\affiliation{$^{1}$Institute of Physics, University of P\'ecs, H-7624 P\'ecs, Hungary}
\affiliation{$^{2}$BME-MTA Exotic Quantum Phases ``Lend\"ulet" Research Group, Budapest Univ. of Technology and Economics,
Institute of Physics, Budafoki \'ut 8., H-1111 Budapest, Hungary}
\date{\today}

\begin{abstract}
If bilayer graphene is placed in a strong perpendicular magnetic field, several quantum Hall plateaus are observed at low enough temperatures.
Of these, the $\sigma_{xy}=4ne^2/h$ sequence ($n\neq0$) is explained by standard Landau quantization, while the other integer plateaus arise due to interactions.
The low-energy excitations in both cases are magnetoexcitons, whose dispersion relation depends on single- and many-body effects in a complicated manner.
Analyzing the magnetoexciton modes in bilayer graphene, we find that the mixing of different Landau level transitions not only renormalizes them,
but essentially changes their spectra and orbital character at finite wave length.
These predictions can be probed in inelastic light scattering experiments.
\end{abstract}


\maketitle

\section{Introduction}

Bilayer graphene,\cite{Novoselov} two coupled hexagonal lattices of carbon atoms in the Bernal stacking\cite{Bernal} of graphite,
is a two-dimensional zero-gap semiconductor with chiral charge carriers with Berry's phase $2\pi$, having a roughly parabolic dispersion at low energies
about the corners of the hexagonal first Brillouin zone.\cite{review}
These facts are testified by its unusual integer quantum Hall effect\cite{Klitzing,qhereview} (IQHE),
featuring a double step in the ladder of the Hall conductance in a strong perpendicular magnetic field $B_\perp$, observed by Novoselov \textit{et al.}\cite{Novoselov}
This double step, $8e^2/h$ instead of the common $4e^2/h$ for spin and valley degenerate Landau levels,
is due to the degeneracy of the $n=0,1$ Landau orbitals.\cite{tightbinding,Pereira,magnetoelectric}
The gap at the integer quantum Hall effect at filling factor $\nu=\rho hc/eB_\perp=\pm4,\pm8$ has been recently measured with great accuracy,\cite{Kurganova}
and the excitations of the IQHE states in the long wavelength limit have also been observed by infrared absorption\cite{Henriksen1} and Raman spectrocopy.\cite{ramanexp}
Further broken symmetry states have been observed\cite{Feldman,Zhao,Weitz,Elferen,Martin} in the central Landau band at $\nu=0,\pm1,\pm2$ and $\pm3$,
and by careful tilted-field measurements it has been shown that they arise predominantly from
many-body effects, i.e., from quantum Hall ferromagnetism (QHF).\cite{Barlas}
Quantum Hall states with broken symmetry have also been found in the $n=-2$ Landau level\cite{Bao}, and there is also some evidence for a fractional quantum Hall plateau.

The eightfold degeneracy of the central group of Landau levels is at best approximate, because the Zeeman energy is unavoidably present.
While the latter is rather small on the characteristic scale of the interaction energy, a perpendicular electric field can be applied to
bias the two layers,\cite{McCann,Ohta,Castro,Min,Oostinga,Kuzmenko,Zhang,Mak,Mucha,Wang,magnetoelectric,dualgate1,dualgate2,dualgate3,Kim,Weitz}
which causes an energy difference between the two valleys.
The competition of the on-site energy difference between the layers and interactions may result in interesting physics, especially at $\nu=0$.\cite{Kim,Weitz,Gorbar,Nand,TF,Kharitonov}

If the chemical potential is in the gap between Landau bands, the low-energy excitations are bound particle-hole pairs,\cite{Kallin,MacD,Lerner,Bychkov} called magnetoexcitons.
As the net charge of such an excitation is zero, taking appropriate linear combinations one obtains eigenstates of the total momentum.
In such states the hole and the particle are bound by the attractive Coulomb interaction,
forming a dipole with a separation of $q\ell_B^2$ at center-of-mass wave-vector $\mathbf q$, where $\ell_B=\sqrt{\hbar/eB}$ is the magnetic length.
These modes determine the transport gap in the $q\to\infty$ limit.
Some of these modes couple to circularly polarized light,\cite{infrared} while others may be observable in
inelastic light scattering experiments.\cite{Pinczuk,Pinczuk3,Pinczuk4,Pinczuk2,Pinczuk6,Pinczuk5,raman,ramanexp}

For monolayer graphene Yang \textit{et al.}\cite{Yang} studied the intra-Landau level excitations of quantum Hall ferromagnetic states,
and Iyengar \textit{et al.},\cite{Iyengar} Bychkov and Martinez,\cite{Bychkov2} Rold\'an \textit{et al.},\cite{Roldan}
and Lozovik and Sokolik\cite{Lozovik} discussed the inter-Landau level excitations in detail both for the IQHE and QHF states.
In particular, the linear dispersion of electrons and holes in graphene makes Kohn's theorem\cite{Kohn} inapplicable.\cite{Roldan}
Many-body corrections to cyclotron resonance due to the interaction with the Dirac sea were calculated by Shizuya.\cite{Shizuya2}
Infrared absorption data by Jiang \textit{et al.}\cite{Jiang} and Henriksen \textit{et al.}\cite{Henriksen2} indicate
a contribution from many-body effects to the inter-Landau level transitions in these states.

In bilayer graphene Henriksen\textit{et al.}\cite{Henriksen1} have found that fitting the single-body parameters does not fully explain the observed cyclotron resonance;
Deacon \textit{et al.}\cite{Deacon} and Zou\textit{et al.}\cite{Zhu} have found a significant particle-hole asymmetry, whose origin is still debated.\cite{Shizuya2}
The wave-vector dependence of the excitations have not been observed so far.
Theoretically, the intra-Landau level excitations in bilayer graphene are well understood.
At odd integers in the central Landau band ($\nu=\pm1,\pm3$) the degeneracy of the $n=0,1$ Landau orbitals causes fluctuations with an in-plane electric dipole character,
which gives rise to unusual collective modes.\cite{Barlas,Barlas2,Cote}
At even integers in the central Landau band ($\nu=\pm2,0$) orbital degeneracy does not play a similar role, and the intra-level excitations are still magnetoexcitons.\cite{Shizuya1,TF}
For inter-LL excitations, the many-body corrections to cyclotron resonance has been calculated by renormalization\cite{Shizuya2}
including the possible particle-hole symmetry-breaking terms but using the unscreened Coulomb interaction, with partial agreement with experiments.\cite{Henriksen1}
To complement these studies, here we address the issue of the inter-Landau level excitations of bilayer graphene in the quantum Hall regime.
We incorporate the screening of the interaction by Landau level mixing, i.e., the interaction-induced mixing of excitonic excitations between different Landau level pairs.
We study the finite wave vector behavior of excitations.

Our paper is organized as follows. In Sec.~\ref{LLsec} we review the tight-binding model of bilayer graphene, and the basic facts concerning its Landau levels and orbitals.
Our goal is to systematically explore the range of applicability of subsequent simplified models,
which neglect several parameters of the Slonczewski-Weiss-McClure\cite{SWM} (SWM) model, or account for them on the level of perturbation theory.
We intend to add a few observations to the excellent studies available in the literature.\cite{tightbinding,Pereira,magnetoelectric}
In Sec.~\ref{kohnnote} we briefly comment on the reasons why Kohn's theorem\cite{Kohn} does not apply for bilayer graphene.
In Sec.~\ref{magnetoexcitons} we review the adaptation of the mean-field theory of magnetoexcitons to the case of bilayer graphene.
In Sec.~\ref{secfilled} we study the excitations of the IQHE states, and in Sec.~\ref{secpartial} those of the QHF states.
We conclude in Sec.~\ref{conclusion}, with an outlook on experimental connections.

\section{Landau levels and orbitals}
\label{LLsec}

\begin{figure}[htbp]
\begin{center}
\includegraphics[width=0.7\columnwidth, keepaspectratio]{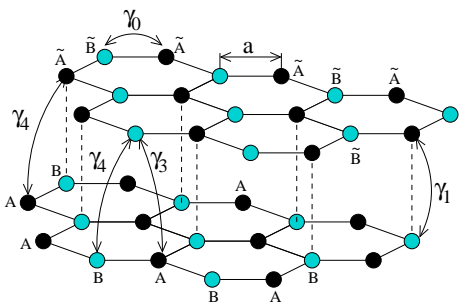}
\end{center}
\caption{\label{stacking}
(Color online)
Bilayer graphene in Bernal stacking. The hopping parameters of the Slonczewski-Weiss-McClure model,\cite{SWM}
conventionally denoted $\gamma_0$, $\gamma_1$, $\gamma_3$ and $\gamma_4$, are also indicated.
}
\end{figure}

Each layer of bilayer graphene consists of two sublattices, denoted $A$ and $B$ in the top layer and $\widetilde A$ and $\widetilde B$ in the bottom layer.
In Bernal stacking\cite{Bernal} two sublattices, $\widetilde A$ and $B$ in our notation, are exactly above/below one another,
while the $A$ sites are above the center of the hexagons in the bottom
layer, and $\widetilde B$ sites are below the centers of hexagons in the top layer. See Fig.~\ref{stacking}.

The low-energy physics of bilayer graphene can be adequately described by the tight-binding effective theories that specialize
the SWM model\cite{SWM} of graphite to the case of just two layers.
In the vicinity of the valley centers corresponding to the $K$ ($\xi=1$) and $K'$ ($\xi=-1$) first Brillouin zone corners,
this amounts to using the Hamiltonian\cite{tightbinding}
\begin{equation}
\label{hamilton}
\hat H_\xi=\xi\begin{pmatrix}
\frac{u-\Delta'}{2} &  v_3\pi & -v_4\pi^\dag &  v\pi^\dag \\
v_3\pi^\dag & -\frac{u+\Delta'}{2} &  v\pi & -v_4\pi \\
-v_4\pi &  v\pi^\dag & -\frac{u-\Delta'}{2} & \xi\gamma_1 \\
v\pi & -v_4\pi^\dag & \xi\gamma_1 & \frac{u+\Delta'}{2}
\end{pmatrix} -\Delta_Z\hat\sigma_z,
\end{equation}
where $\pi=p_x+ip_y$ and $\mathbf p=-i\hbar\nabla -e\mathbf A$, $v=\sqrt3 a\gamma_0/2\hbar\approx 10^6$ m/s is the intra-layer velocity,
$v_3=\sqrt3 a\gamma_3/2\hbar$ is the trigonal warping parameter,
$\gamma_1$ is the inter-layer hopping amplitude, and $v_4=\sqrt3 a\gamma_4/2\hbar$ is a velocity parameter related to interlayer next-nearest neighbor hopping.
$\Delta_Z=g\mu_BB_\perp$ is the Zeeman energy (with $g$ being the gyromagnetic factor and $\mu_B$ the Bohr magneton).
This Hamiltonian acts in the basis of sublattice Bloch states $[\psi_A,\psi_{\widetilde B},\psi_{\widetilde A},\psi_B]$ in valley $K$
and $[\psi_{\widetilde B},\psi_A,\psi_B,\psi_{\widetilde A}]$ in valley $K'$.
Here $\gamma_0=\gamma_{AB}=\gamma_{\widetilde A\widetilde B}$ is the intra-layer hopping amplitude,
$\gamma_1=\gamma_{\widetilde AB}$ is the interlayer hopping amplitude between sites above each other in the two layers.
Further, $\gamma_3=\gamma_{A\widetilde B}$  and $\gamma_4=\gamma_{A\widetilde A}=\gamma_{B\widetilde B}$ are next-nearest neighbor interlayer hopping amplitudes,
as shown in Fig.~\ref{stacking}.
$\Delta'$ is the on-site energy difference between the dimer sites ($\widetilde A,B$) and the non-dimer sites  ($\widetilde B,A$).
Finally, $u$ is the potential energy difference between the layers, which may arise, e.g., because of an applied perpendicular electric field $E_\perp$.

The Hamiltonian in Eq.~(\ref{hamilton}) is block-diagonal in the valley index, which is conveniently described as a pseudospin.
In the special case $u=0$ the system has SU(2) pseudospin rotation symmetry.
In the theoretical limit $\Delta_Z\to0$ this is raised to SU(4) symmetry.
In this paper we will treat $\Delta_Z$ and $u$ as small perturbations in comparison to the interaction energy, i.e., we will work in the
$\Delta_Z,u\ll e^2/(4\pi\epsilon_0\epsilon_r\ell_B)$ limit, where $\epsilon_r$ is the relative dielectric constant of the environment.
We set $\hbar=1$.

For small momenta, $p\ll\gamma_1/4v$, the two low-energy bands of the Hamiltonian $\hat H_\xi$ that touch each other at $K$ and $K'$ 
in the case of vanishing magnetic field can be attributed to a $2\times2$ effective Hamiltonian\cite{McCann}
\begin{multline}
\label{lowenergy}
\hat H'_\xi=-\frac{1}{2m}\begin{pmatrix}
0 & (\pi^\dag)^2 \\ \pi^2 & 0
\end{pmatrix}+
\xi v_3\begin{pmatrix}
0 & \pi \\ \pi^\dag & 0
\end{pmatrix}+\\
+\xi u\left(
\frac{1}{2}\begin{pmatrix}
1 & 0\\0 & -1\end{pmatrix}
-\frac{v^2}{\gamma_1^2}\begin{pmatrix}
\pi^\dag\pi & 0\\0 & -\pi\pi^\dag\end{pmatrix}
\right)
\end{multline}
where $m=\gamma_1/2v^2$, and $\hat H'_\xi$ acts on $[\psi_A,\psi_{\widetilde B}]$ in valley $K$ and $[\psi_{\widetilde B},\psi_A]$ in valley $K'$.
The Landau levels (LL's) and Landau orbitals, respectively, of the two-band Hamiltonian $\hat H'_\xi$ are
\begin{gather}
E_{0\xi}=\frac{\xi u}{2},\quad\quad
E_{1\xi}=\frac{\xi u}{2}-\xi\frac{u\hbar\omega_c}{\gamma_1},\\
E_{n\xi}=\text{sgn}(n)\hbar\omega_c\sqrt{|n|(|n|-1)}-\xi\frac{u\hbar\omega_c}{2\gamma_1},\\
\Psi_{0\text{ or }1,q}=\begin{pmatrix} \eta_{0\text{ or }1,q} \\ 0 \end{pmatrix},\quad
\Psi_{nq\xi}=\begin{pmatrix} A^{(n)}_\xi\eta_{|n|q} \\ B^{(n)}_\xi\eta_{|n|-2,q} \end{pmatrix},\label{genericorbitals}\\
A^{(n)}_\xi=\frac{1}{C^{(n)}_\xi},\quad
B^{(n)}_\xi=\frac{1}{C^{(n)}_\xi}\frac{E_{n\xi}-\xi\frac{u}{2}+\xi\frac{u|n|\hbar\omega_c}{\gamma_1}}{\hbar\omega_c\sqrt{|n|(|n|-1)}}.\nonumber
\end{gather}
Here $n\neq-1$ is an integer, $C^{(n)}_\xi$ is an appropriate normalization factor,
and $\eta_{nq}$ are the single-particle states in the conventional two-dimensional electron gas (2DEG) with quadratic dispersion in the Landau gauge $\mathbf A=\mathbf{\hat y}Bx$,
\begin{equation}
\label{etaeq}
\eta_{nq}(\mathbf r)=\frac{e^{iqx-\left(y/\ell_B-q\ell_B\right)^2/2}}{\sqrt{2\pi\sqrt\pi 2^n n!\ell_B}}H_n\left( \frac{y}{\ell_B}-q\ell_B \right),
\end{equation}
and $H_n$ is a Hermite-polynomial.

The $n=0,1$ orbitals are degenerate in the $u\to0$ limit, and they have a layer polarization for $\xi=\pm1$.
At realistic values of $\Delta_Z$ and $u$, the $n=0,1$, $\xi=\pm1$,
$\sigma=\uparrow,\downarrow$ states form a quasidegenerate band we will call the central Landau level octet.
Notice that $A^{(n)}_\xi\to1/\sqrt2$ and $|B^{(n)}_\xi|\to1/\sqrt2$ for $u\to0$.

We would like to determine how neglecting $\gamma_3$, $\gamma_4$, and $\Delta'$ changes the single-body orbitals of the four-band Hamiltonian $\hat H_\xi$ in Eq.~(\ref{hamilton}),
and how much these differ from the simplified two-band model $\hat H'_\xi$ in Eq.~(\ref{lowenergy}).
The values of the SWM parameters for bilayer graphene was estimated by a combination of infrared response analysis and
theoretical techniques by Zhang \textit{et al.}\cite{Zhang2} They found
\begin{equation}
\frac{\gamma_1}{\gamma_0}=0.133,\label{estim}\quad
\frac{\gamma_3}{\gamma_0}=0.1,\quad
\frac{\Delta'}{\gamma_0}=0.006.
\end{equation}
These ratios are based on $\gamma_0=3.0$ eV.
While somewhat greater values of $\gamma_0$ are also available in the literature,\cite{Li} we use these values for a robustness analysis.
For the particle-hole symmetry breaking term we use
\begin{equation}
\frac{\gamma_4}{\gamma_0}=0.063\label{estimw4}
\end{equation}
from the recent electron and hole mass measurement by Zou \textit{et al.},\cite{Zhu} which is slightly greater than the value in Ref.~\onlinecite{Zhang2}.

\begin{figure}[htbp]
\begin{center}
\includegraphics[width=\columnwidth, keepaspectratio]{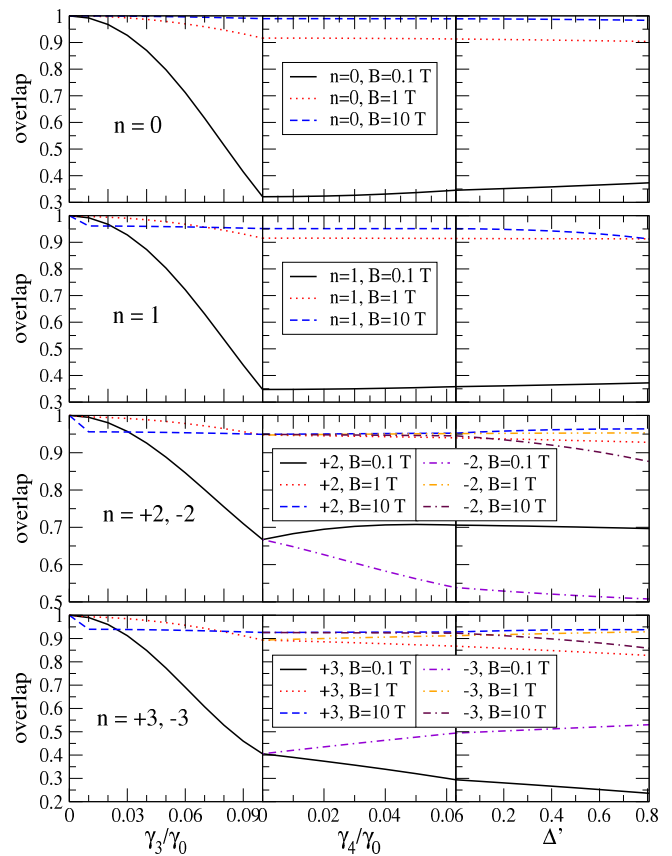}
\end{center}
\caption{\label{overlap}
(Color online)
The overlap of the Landau orbitals with the ``ideal'' limit, $\gamma_3=\gamma_4=\Delta'=0$, as the SWM parameters $\gamma_3,\gamma_4,\Delta'$ are gradually tuned from zero to their literary values
[Eqs. (\ref{estim}) to (\ref{estimw4})] for the lowest-energy Landau levels.
For the effect of $\gamma_3$ in the two-band model see Ref.~\onlinecite{Misumi}.
}
\end{figure}

With $\pi=\frac{\sqrt2\hbar}{i\ell_B}a$, $[a,a^\dag]=1$, the Hamiltonian can be expressed in terms of these Landau level ladder operators.
Then the eigenstates of $\hat H_\xi$ can be calculated numerically.
Fig.~\ref{overlap} shows the overlap of the Landau orbitals with the ``ideal'' limit $\gamma_3=\gamma_4=\Delta'=0$ as the SWM parameters $\gamma_3,\gamma_4,\Delta'$
are tuned from zero to their literary values [Eqs. (\ref{estim}) to (\ref{estimw4})] for the central ($n=0,1$) and the two pairs of nearby ($n=\pm2,\pm3$) Landau levels.
At small magnetic field $B=0.1$ T trigonal warping alone significantly changes the orbitals from their ideal limit.
Switching on $\gamma_4$  hardly affects the central levels, but for $n\ge2$ it changes the electron and hole pairs ($+n,-n$) differently,
as expected from this electron-hole symmetry breaking term.
Finally, the inclusion of $\Delta'$ hardly affects the orbitals.
These changes, however, are already small at modest fields ($B=1$ T), and are further suppressed as the experimentally relevant range ($B\approx10$ T) is approached.
Thus neglecting the $\gamma_3,\gamma_4,\Delta'$ SWM parameters is justified in the high magnetic field range where quantum Hall experiments are typically performed.

As the two-band model in Eq.~(\ref{lowenergy}) applies for small momenta, and the low-index Landau orbitals have a small amplitude at high momenta,
the two-band model is expected to be valid for the lowest few Landau levels.
The Landau orbitals of the two-band model have a large overlap with those of the four-band model in the  ``ideal'' limit $\gamma_3=\gamma_4=\Delta'=0$:
1, 0.9995, 0.9992, 0.9987 for $n=0,1,\pm2,\pm3$, respectively.
We conclude that using the Landau states of two-band model in Eq.~(\ref{lowenergy}) instead of these of the
four-band model $\hat H_\xi$ in the lowest-energy Landau bands does not introduce further inaccuracy beyond the neglect of $\gamma_3,\gamma_4$ and $\Delta'$.
Therefore, we take $\hat H'_\xi$ as our starting point.

\section{A note on Kohn's theorem}
\label{kohnnote}

Kohn's theorem\cite{Kohn} states that interactions do not shift the cyclotron resonance in a parabolic band.
It applies equally to two- and three-dimensional systems.
It is not applicable to linear bands in monolayer graphene.\cite{Roldan}
We will see that it also fails for bilayer graphene, even though the bands of $\hat H_\xi$ in Eq.~(\ref{hamilton}) start quadratically at low energies,
and those of $\hat H'_\xi$ in Eq.~(\ref{lowenergy}) are exactly parabolic for $v_3=u=0$.

For the conventional 2DEG, Kohn's theorem follows because the interaction with a radiation field,
\[
\hat H_{\text{EM}}=\sum_i\frac{\mathbf p_i\cdot\delta\mathbf A(\mathbf r_i)}{m},
\]
where $\mathbf p_i$ is the canonical momentum that includes vector potential of the homogeneous magnetic field, is proportional to
$P = P_x+iP_y$, where $P_x=\sum_i p_{i,x}$ and $P_y=\sum_i p_{i,y}$.
$P_x$ and $P_y$ are generators of global translations, hence they commute with any translation-invariant interaction;
moreover, $P,P^\dag$ act as ladder operators among the eigenstates of the total kinetic energy, $\hat H_0=\sum_i\mathbf p_i^2/2m$.
Therefore, $\hat H_{\text{EM}}$ connects eigenstates of the total Hamiltonian and conserves the interparticle interaction.
$\delta\mathbf A$ may be described classically or quantum mechanically; $[\hat H_{\text{EM}},\hat V]=0$ can also be checked directly.

In the two-band model of bilayer graphene the interaction with the radiation field $\delta\mathbf A$ is\cite{infrared,raman}
\[
\hat H_{\text{EM}}=\frac{e}{m}\sum_i\begin{pmatrix}
0 && \delta A^\dag(\mathbf r_i)\pi_i^\dag \\
\delta A(\mathbf r_i)\pi_i && 0
\end{pmatrix}
\]
with $\delta A=\delta A_x+i\delta A_y$.
It is straightforward to show that $\hat H_{\text{EM}}$ maps a state $\Psi_{nq}$ to a linear combination
of $\Psi_{n+1,q'}$, $\Psi_{n-1,q'}$, $\Psi_{-n-1,q'}$ and $\Psi_{-n+1,q'}$.
If three out of these transitions are Pauli-blocked, we may end up in an eigenstate of the kinetic part of the many-body Hamiltonian,
but $\hat H_{\text{EM}}$ is no longer proportional to a linear combination of $P_x$ and $P_y$.
In fact, it no longer commutes with them.
Thus the interaction energy may differ in the electromagnetically excited many-body state and the initial state.

\section{Magnetoexcitons}
\label{magnetoexcitons}

When the Fermi energy is in a Landau gap (e.g. at filling factor $\nu=\dots,-12,-8,-4,4,8,12,\dots$ in bilayer graphene), 
the integer quantum Hall effect\cite{Klitzing} occurs in samples with moderate disorder.\cite{Novoselov}
Quantum Hall states also occur at other integer filling factors because the exchange interaction favors symmetry-breaking ground states
called quantum Hall ferromagnets; single-body terms such as the Zeeman energy play a secondary role.
QHF's emerge at odd integer fillings in two-component systems even if the Zeeman energy is tuned to zero.
This observation straightforwardly generalizes for SU($n$) systems.\cite{Yang}

Because of the clear separation of the filled and empty Landau bands in the (mean-field) ground-state,
the excitations of both classes of quantum Hall systems are described in the same way.
The relevant low-energy excitations are magnetoexcitons,\cite{foot2}
which are obtained by promoting an electron from a filled Landau band to an empty band.\cite{Kallin,MacD,Lerner,Bychkov}
These neutral excitations have a well-defined center-of-mass momentum $\mathbf Q$.
They approach widely separated particle-hole pairs in the $Q\to\infty$ limit.
The latter limit determines the transport gap unless skyrmions form.\cite{skyrmions}
Magnetoexcitons are created from the ground-state by operators\cite{Kallin,MacD,Lerner,Bychkov}
\begin{equation}
\label{exciton}
\hat\Psi^\dag_{NN'}(\mathbf Q)=\sqrt{\frac{2\pi\ell_B^2}{A}}\sum_p e^{ipQ_y\ell_B^2}\hat a^\dag_{Np}\hat a_{N'p-Q_x},
\end{equation}
where $N=(n,\xi,\sigma)$ ($N'=(n',\xi',\sigma')$) specifies the Landau band where the particle (hole) is created and $A$ is the area of the sample.

Magnetoexcitons carry spin and pseudospin (valley) quantum numbers, as derived from the particle and hole Landau bands involved.
While the projections $S_z,P_z$ of the spin and the pseudospin are always good quantum numbers,
their magnitudes $S$ and $P$ are well-defined only for ground-states that are spin or pseudospin singlets, respectively.

It is common practice\cite{Kallin,Iyengar} to define the quantity
\begin{equation}
l_z=|n|-|n'|,\label{approx}
\end{equation}
and consider it the ``angular momentum quantum number'' of the exciton.
We emphasize that $l_z$ is exactly conserved by the electron-electron interaction only in the $\mathbf Q\to0$ limit, where it is related to angular momentum.
The emergence of this quantity is best seen in the two-body problem of the negatively charged electron and the positively charged hole, as discussed in the Appendix.
At any finite wave vector transitions with different $\l_z$ may mix.

In the low magnetoexciton density limit the interaction between magnetoexcitons is neglected.
The mean-field (Hartree-Fock) Hamiltonian of magnetoexcitons is well known from the literature,\cite{Kallin,MacD,Lerner,Bychkov}
and so is its adaptation to spinorial orbitals:\cite{Iyengar,Bychkov2,TF}
\begin{multline}
H^{(\tilde N\tilde N')}_{(NN')}(\mathbf Q)=\\
\langle 0|\Psi_{\tilde N\tilde N'}(\mathbf Q)\hat V\Psi^\dagger_{NN'}(\mathbf Q) |0\rangle-\delta_{N\tilde N}\delta_{N'\tilde N'}\langle0|\hat V|0\rangle=\\
=\delta_{N\tilde N}\delta_{N'\tilde N'}\left( E_{n\alpha\xi\sigma}-E_{n'\alpha'\xi'\sigma'} + \Delta(n,n')\right)+\\
+E^{(\tilde N\tilde N')}_{(NN')}(\mathbf Q)
+R^{(\tilde N\tilde N')}_{(NN')}(\mathbf Q),\label{Hmf}
\end{multline}
where $N=(n,\xi,\sigma)$, etc., and $\delta_{NN'}=\delta_{\sigma\sigma'}\delta_{\xi\xi'}\delta_{nn'}$.
The first term of the r.h.s.\ is the single-body energy difference of the $N$ and $N'$ states,
which includes the wave vector independent exchange self-energy difference of the two states.
While the exchange self-energy itself is infinite for any orbital, its difference between two states,
\begin{align}
X_{N'N}&=\int\frac{d\mathbf q}{(2\pi)^2}I^{N'N}_{N'N}(\mathbf p),\\
\Delta(N,N')&=\sum_{M\text{ filled}}\left(X_{N'M} - X_{MN}\right),
\end{align}
is finite. (We will define $I^{N_2N_2'}_{N_1N_1'}(\mathbf p)$ soon.)
This is a peculiarity of bilayer graphene.
A simple regularization procedure works for the four-band model.\cite{Shizuya3}
For monolayer graphene, a proper renormalization procedure is required.\cite{Iyengar,Shizuya2}
The shift $\Delta(N,N')$ is analogous to the Lamb shift in quantum electrodynamics.\cite{Shizuya3}

The next term is the direct dynamical interaction between the electron and the hole:
\begin{equation}
E^{(\tilde N\tilde N')}_{(NN')}(\mathbf Q)=-\int\frac{d\mathbf q}{(2\pi)^2}e^{i\mathbf{\hat z}\cdot(\mathbf q\times\mathbf Q)} I^{N\tilde N}_{N'\tilde N'}(\mathbf q).
\end{equation}
This term is diagonal both in spin and pseudospin, $\propto\delta_{\tilde\sigma\sigma}\delta_{\tilde\xi\xi}\delta_{\tilde\sigma'\sigma'}\delta_{\tilde\xi'\xi'}$,
but not in Landau orbital indices.
Finally, the last term in Eq.~(\ref{Hmf}) is the exchange interaction between the electron and the hole,
\begin{equation}
R^{(\tilde N\tilde N')}_{(NN')}(\mathbf Q)=\frac{1}{2\pi\ell_B^2}\mathfrak{Re}I^{NN'}_{\tilde N\tilde N'}(\mathbf Q),
\end{equation}
which is $\propto\delta_{\sigma\sigma'}\delta_{\xi\xi'}\delta_{\tilde\sigma\tilde\sigma'}\delta_{\tilde\xi\tilde\xi'}$,
thus couples transitions that conserve the spin $\sigma$ and the valley $\xi$ of the electron and the hole individually.
Sometimes we will call it the RPA contribution, as it is related to particle-hole annihilation and recreation processes.
(It is also called\cite{Roldan} the depolarization term.)
Notice $R^{(\tilde N\tilde N')}_{(NN')}(\mathbf Q)$ vanishes in the $Q\to0$ limit.

We have used the notation
\begin{widetext}
\begin{gather*}
I^{N_2N_2'}_{N_1N_1'}(\mathbf p)=V^S(p)\left[
A^{(n_2')}_{\xi_2'} A^{(n_1)}_{\xi_1} A^{(n_1')}_{\xi_1'} A^{(n_2)}_{\xi_2} F^\ast_{|N_2||N_2'|}(\mathbf p)F_{|N_1||N_1'|}(\mathbf p)+
\right.\\
\left.
B^{(n_2')}_{\xi_2'} B^{(n_1)}_{\xi_1} B^{(n_1')}_{\xi_1'} B^{(n_2)}_{\xi_2} F^\ast_{|N_2|-2,|N_2'|-2,}(\mathbf p)F_{|N_1|-2,|N_1'|-2}(\mathbf p)\right]+\\
V^D(p)\left[
A^{(n_2')}_{\xi_2'} B^{(n_1)}_{\xi_1} B^{(n_1')}_{\xi_1'} A^{(n_2)}_{\xi_2} F^\ast_{|N_2||N_2'|}(\mathbf p)F_{|N_1|-2,|N_1'|-2}(\mathbf p)+
B^{(n_2')}_{\xi_2'} A^{(n_1)}_{\xi_1} A^{(n_1')}_{\xi_1'} B^{(n_2)}_{\xi_2} F^\ast_{|N_2|-2,|N_2'|-2,}(\mathbf p)F_{|N_1||N_1'|}(\mathbf p)\right],
\end{gather*}
\[
F_{N'N}(\mathbf q)=\delta_{\sigma\sigma'}\delta_{\xi\xi'}\sqrt\frac{n!}{(n')!}
\left(\frac{(-q_y+iq_x)\ell_B}{\sqrt2}\right)^{n'-n}L_n^{n'-n}\left(\frac{q^2\ell_B^2}{2}\right)e^{-q^2\ell_B^2/4}\quad
\text{if $n'\ge n$, else } F_{NN'}(\mathbf q)=F^\ast_{N'N}(-\mathbf q).
\]
\end{widetext}
Here $N-2\equiv(n-2,\xi\sigma)$, $|N|=(|n|,\xi,\sigma)$, and $F_{N'N}(\mathbf q)$ is related to the Fourier transform of $\eta_{nq}(\mathbf r)$ in Eq.~(\ref{etaeq}),
and $L^m_n(z)$ is an associated Laguerre polynomial.
The difference between the intralayer Coulomb interaction $V^S(q)=2\pi e^2/(4\pi\epsilon_0\epsilon_r q)$ and the
intralayer one, $V^D(q)=e^{-qd}V^S(q)$ where $d\approx0.335$ nm is the distance between the layers, is neglected as a first approximation.
The $A^{(n)}_\xi$ and $B^{(n)}_\xi$ numbers correspond to the spinorial structure of the single-body states in Eq.~(\ref{genericorbitals}):
$A^1=A^0=1$, $B^1=B^0=0$ and $A^{(n)}_\xi=\text{sgn}(n)B^{(n)}_\xi=1/\sqrt2$ for $n\ge2$.
Notice that 
\begin{gather}
E^{(\tilde N\tilde N')}_{(NN')}(\mathbf Q)=E^{(NN')}_{(\tilde N\tilde N')}(\mathbf Q)=E^{(\tilde N'\tilde N)}_{(N'N)}(\mathbf Q),\\
R^{(\tilde N\tilde N')}_{(NN')}(\mathbf Q)=R^{(NN')}_{(\tilde N\tilde N')}(\mathbf Q)=(-1)^{n+n'+\tilde n+\tilde n'}R^{(\tilde N'\tilde N)}_{(N'N)}(\mathbf Q),
\end{gather}
which follow from the similar properties of $I^{N_2N_2'}_{N_1N_1'}(\mathbf p)$.

The $q\to0$ limit of the magnetoexciton dispersion determines the many-body contribution to the cyclotron resonance,
which may be nonvanishing in graphene systems (c.f.\ Sec.~\ref{kohnnote} and Refs.~\onlinecite{Roldan} and \onlinecite{Shizuya2}).

The mean field Hamiltonian matrix $H^{(\tilde N\tilde N')}_{(NN')}(\mathbf Q)$ in general mixes transitions among different electron-hole pairs, restricted only by conservation laws.
Landau level mixing effectively screens the interaction.
Sometimes the magnetoexciton spectra are obtained using a screened model interaction instead of the bare Coulomb, not letting LL transitions mix.\cite{Misumi,Shizuya2}
We believe such an approach is suitable in the $q=0$ limit, where an additional quantum number $l_z$ also restricts LL mixing, and for intra-LL modes.
At finite wave vector the mean-field theory with LL mixing removes spurious level crossings in the excitation spectra and provides insight into the orbital structure of the excitations.
Technically, however, the infinite $H^{(\tilde N\tilde N')}_{(NN')}(\mathbf Q)$ matrix needs to be truncated.


\section{Integer quantum Hall states}
\label{secfilled}

We first consider the states where the chemical potential is between two orbital Landau bands.
This occurs at filling factor $\nu=\dots,-12,-8,-4,4,8,12,\dots$ in bilayer graphene.
Together with $S_z$ and $P_z$, the magnitude of the spin $S$ and of the pseudospin $P$ are quantum numbers.
(In the $\Delta_Z\to0,u\to0$ limit an SU(4) classification is also possible.)

\begin{figure}[htbp]
\begin{center}
\includegraphics[width=\columnwidth, angle=0, keepaspectratio]{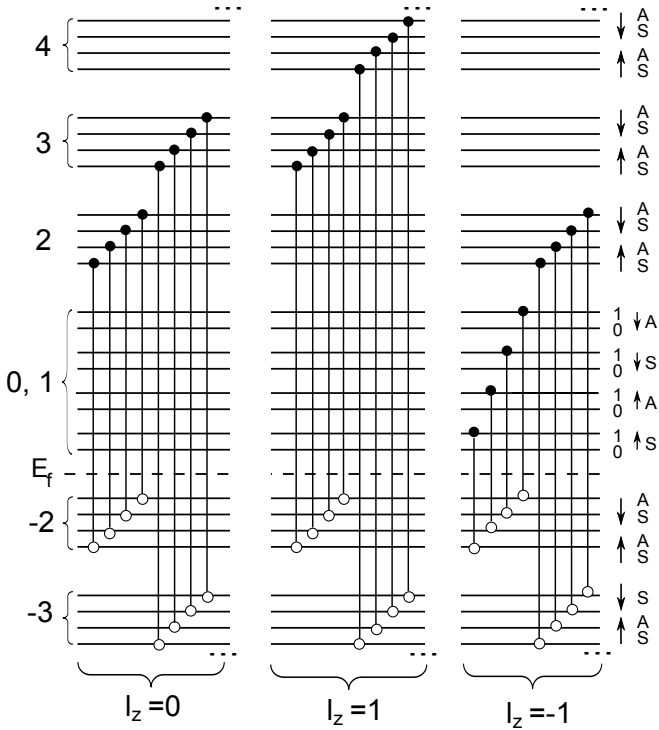}
\end{center}
\caption{\label{sketchinteger}
Magnetoexciton modes at the $\nu=-4$ integer quantum Hall effect in bilayer graphene.
Only the optically relevant spin- and pseudospin-conserving modes are shown.
In the $q\to0$ limit, which is probed by purely optical experiments, the transitions in the infinite sequence at fixed $l_z=|n|-|n'|$ may mix;
at $q>0$, the excitations result from the mixing of all $l_z$ sequences.
The modes at $\nu=+4$ are obtained by particle-hole conjugation.
}
\end{figure}

With the hole ($n'$) and the electron ($n$) Landau orbitals fixed, the sixteen possible transitions belong to four classes:
(i) a spin singlet, pseudospin singlet state:
\begin{equation}
\label{m0}
\hat\Psi^{\dag00}_{nn'}(\mathbf Q)=\frac{1}{2}\sum_\xi\sum_\sigma\hat\Psi^\dag_{n\sigma\xi,n'\sigma\xi}(\mathbf Q).
\end{equation}
(ii) A spin singlet, pseudospin triplet multiplet. The $P_z=0$ member of this multiplet is
\begin{equation}
\hat\Psi^{\dag01}_{nn'}(\mathbf Q)=\frac{1}{2}\sum_\xi\sum_\sigma\text{sgn}(\xi)   \hat\Psi^\dag_{n\sigma\xi,n'\sigma\xi}(\mathbf Q).
\end{equation}
(iii) A spin triplet, pseudospin singlet, which contains following the $S_z=0$ state:
\begin{equation}
\hat\Psi^{\dag10}_{nn'}(\mathbf Q)=\frac{1}{2}\sum_\xi\sum_\sigma\text{sgn}(\sigma)   \hat\Psi^\dag_{n\sigma\xi,n'\sigma\xi}(\mathbf Q).
\end{equation}
(iv) A nine-member multiplet that is triplet is both spin and valley. Its $S_z=0=P_z$ member is
\begin{equation}
\hat\Psi^{\dag11}_{nn'}(\mathbf Q)=\frac{1}{2}\sum_\xi\sum_\sigma\text{sgn}(\sigma)\text{sgn}(\xi)\hat\Psi^\dag_{n\sigma\xi,n'\sigma\xi}(\mathbf Q).
\end{equation}
Fig.~\ref{sketchinteger} depicts these modes for $\nu=-4$.

The exchange interaction between the electron and the hole contributes only to states generated by $\{\hat\Psi^{\dag00}_{nn'}\}_{nn'}$.
In all other excitation modes the RPA term cancels (c.f.\ the sign alternation\cite{foot} in modes $\hat\Psi^{\dag01}_{nn'}$, $\hat\Psi^{\dag10}_{nn'}$ and $\hat\Psi^{\dag11}_{nn'}$)
or is prohibited by quantum numbers.
Thus in the absence of accidental degeneracies, we expect a collection of nondegenerate excitations and one of fifteen-fold degenerate excitations;
the latter is decomposed as $4,7,4$ if $\Delta_Z>0,u=0$ and as $1,2,1,2,3,2,1,2,1$ if $\Delta_Z,u>0$.

The extent interactions may mix transitions involving different Landau level pairs depends on the interaction-to-kinetic energy ratio, parametrized by
\begin{equation}
\beta=\frac{e^2}{4\pi\epsilon_0\epsilon_r\ell_B}\Big/\hbar\omega_c\propto\frac{1}{\epsilon_r\sqrt B}.
\end{equation}
Notice that $\beta\to0$ in the $B\to\infty$ limit just like for the conventional two-dimensional electron gas.
Realistically ($10\text{ T}\le B\le40$ T, $1\le\epsilon_r\le4$), $1<\beta<8$; this is by no means a small perturbation.

In the conventional 2DEG LL mixing is suppressed at high fields because of the $B^{-1/2}$ scaling of the relative 
strength of interactions, while in monolayer graphene both the interaction and the kinetic energy scale with $B^{1/2}$, thus LL mixing is never suppressed.
In bilayer graphene the kinetic term of the Hamiltonian interpolates between quadratic at small momenta and linear at high momenta;
thus LL mixing gets suppressed only for the LL's with a small index, whose orbitals are built up from low-momentum plane waves.
For high-index LL's the ratio of interaction to kinetic energy is only weakly $B$-dependent.
At fixed magnetic field and filling factor, LL mixing in bilayer graphene is more significant than in a conventional 2DEG,
implemented, e.g., in GaAs quantum quantum wells, because of the smaller dielectric constant and effective mass in bilayer graphene.

One can also compare the Coulomb energy scale, $e^2/(4\pi\epsilon_0\epsilon_r\ell_B)$ to the energy difference between adjecent Landau levels,
$\hbar\omega_c\left(\sqrt{n(n-1)} - \sqrt{(n-1)(n-2)}\right)$.
This may give the impression that LL mixing is more important at higher filling factors, but the amplitude of the undulations of the
unmixed magnetoexciton dipersions also gets reduced in higher levels; leaving the issue of the generic progress of LL mixing with increasing filling factor open.

\begin{figure*}[htbp]
\begin{center}
\includegraphics[width=\textwidth, angle=0, keepaspectratio]{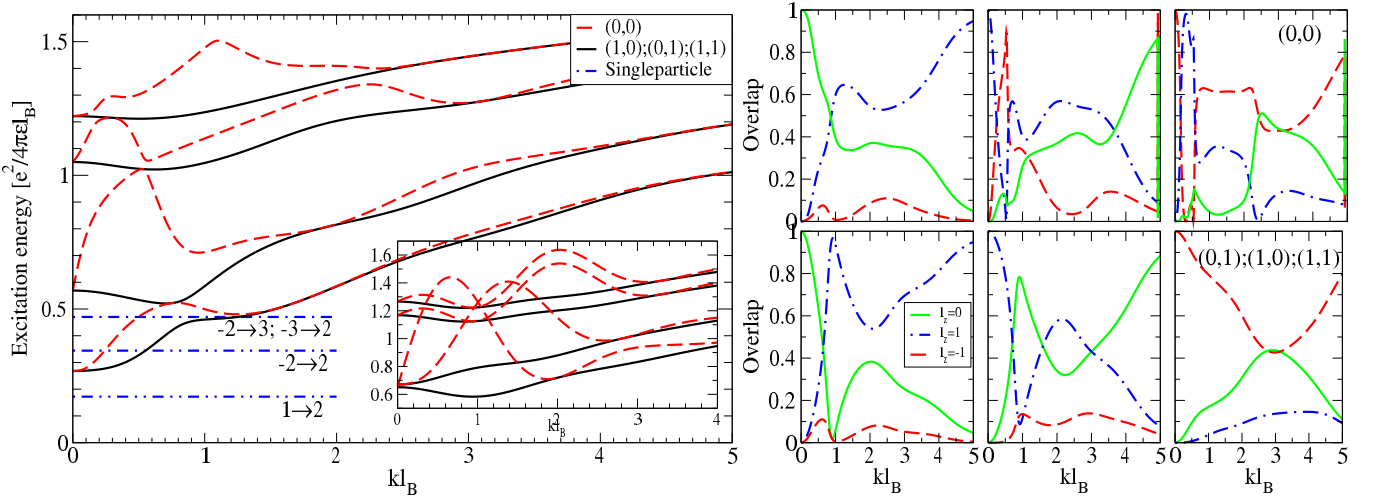}
\end{center}
\caption{\label{integer}
(Color online)
The excitations of the integer quantum Hall state at $|\nu|=4$ and $B=10T$.
The mixing of Landau levels is truncated at $L=1$ and $M=7$.
Solid lines show the fifteen-fold degenerate excitations, which include three optically relevant $S_z=P_z=0$ modes.
Dashed lines show the spin and pseudospin singlets.
Inset: spectra if Landau level mixing is neglected.
Side panels: the weight of the definite $l_z$ projections in each curve in bottom-up order.
The top row shows the spin and pseudospin singlets.
}
\end{figure*}

\begin{figure}[htbp]
\begin{center}
\includegraphics[width=\columnwidth,keepaspectratio]{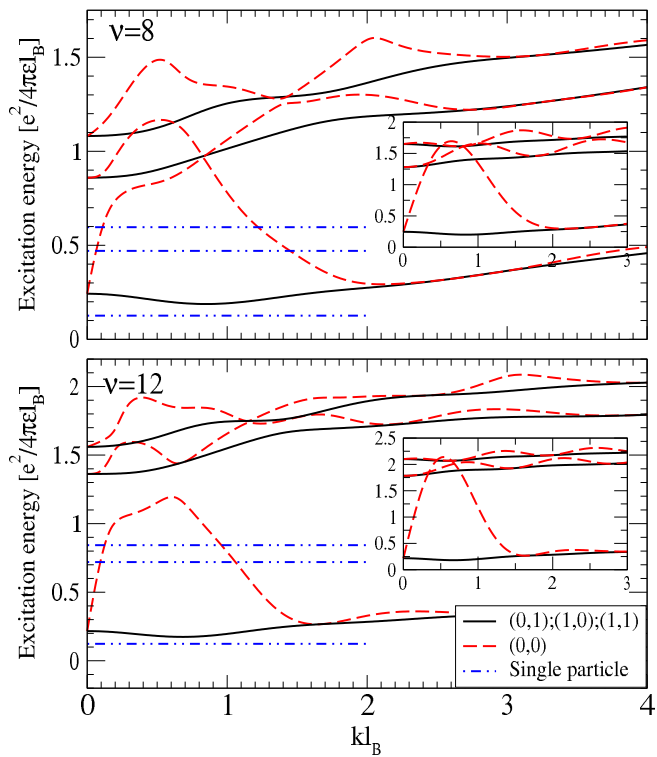}
\end{center}
\caption{\label{integer2}
(Color online)
The excitations of the integer quantum Hall state at $|\nu|=8$ and $|\nu|=12$, at $B=10T$.
}
\end{figure}

For $q=0$ the mean-field Hamiltonian mixes magnetoexcitons with different electron and hole Landau levels at fixed $l_z$,
and for $q>0$ it also mixes different $l_z$ subspaces; c.f.\ Fig.~\ref{sketchinteger}.
Restricting LL mixing to a fixed $l_z$ subspace might give the impression that LL mixing is just a quantitative correction for the long wave length
part of the lowest excitation curves, resulting in increased electron-hole binding energies.\cite{Lozovik,Moska}
However, already the lowest excitations in the different $l_z$ sectors mix strongly at finite wave vector.
As the side panels of Fig.~\ref{integer} show, the excitations have a large projection on the $l_z$ subspaces different from their own $l_z$ in the $q\to0$ limit,
and may eventually be contained in one of the other subspaces for large $q$; this is an unavoidable consequence of the elimination of crossings by LL mixing.
The mixing of Landau levels is especially strong in the nondegenerate excitations, which are strongly affected by the exchange interaction
between the electron and the hole (the RPA term) in the $q\ell_B\sim1$ region.
Thus in the rest of this paper we allow the mixing of transitions restricted by $|l_z|\le L$ and a maximum number $M$ at each fixed $l_z$.
Fig.~\ref{integer} shows transitions at $\nu=4$ with $L=1$ and $M=7$, while Fig.~\ref{integer2} shows $|\nu|=8$ and 12.
We will use this truncation in the spectra shown in the rest of this paper.\cite{foot1}
Figs.~\ref{integer} and \ref{integer2} also show the kinetic energy difference between the electron and the hole for comparison.
Our mean-field theory predicts an interaction shift comparable to this energy.
This prediction will be revisited with methods beyond mean-field.\cite{future}

The $q\to0$ limit of the magnetoexcitons are commonly probed by optical absorption and electronic Raman scattering.
The selection rules\cite{infrared,raman} ensure that only the $\hat\Psi^{\dag00}_{nn'}$ mode and the $\hat\Psi^{\dag01}_{nn'}$, $\hat\Psi^{\dag10}_{nn'}$ and $\hat\Psi^{\dag11}_{nn'}$
modes of the fifteen-fold degenerate curve are active, $l_z=\pm1$ is absorption and $l_z=0$ in Raman.
Particle-hole conjugation relates $\nu=4n$ to $\nu=-4n$ ($n$ integer) with the sign of $l_z$ reversed.

\section{Quantum Hall ferromagnetic states}
\label{secpartial}

With an integer filling factor different from $\nu=\pm 4,\pm8,\pm12,\dots$,
a Landau band quartet ($|\nu|>4$) or octet ($|\nu|<4$) is partially filled in the single electron picture.
The minimization of the interaction energy results in gapped states which break either spin rotation or pseudospin (valley) rotation symmetry,
or both.\cite{Barlas,Barlas2,Cote,Bisti,Yang,Gorbar,TF}
If either the Zeeman energy $\Delta_Z$ or the interlayer energy difference $u$ is present, they affect the order how the Landau levels are filled,
but exchange energy considerations are more crucial in most cases.\cite{Barlas,Shizuya3}
The most convenient basis in pseudospin space may differ; we may introduce
\begin{align}
\hat a_{n S\sigma p}&=\cos\frac{\theta}{2}\hat a_{n,\xi=1,\sigma p}+\sin\frac{\theta}{2}e^{i\phi}\hat a_{n,\xi=-1,\sigma p},\label{defS}\\
\hat a_{n A\sigma p}&=\sin\frac{\theta}{2}\hat a_{n,\xi=1,\sigma p}-\cos\frac{\theta}{2}e^{i\phi}\hat a_{n,\xi=-1,\sigma p}.\label{defA}
\end{align}
With a proper choice of $\theta$ and $\phi$, Eqs.~(\ref{defS}-\ref{defA}) include states of definite valley, bonding and antibonding states, or intervalley phase coherent states.
Corresponding magnetoexciton operators are defined in an obvious manner.

In particular, if $\Delta_Z>u$, the $\nu=0$ QHF state is ferromagnetic and the choice of the pseudospin basis is irrelevant (Fig.~\ref{sketchnu0}).
For $\nu=\pm2$, both the $n=0$ and $n=1$ orbital Landau levels of identical spin and pseudospin are filled, where $\phi$ and $\theta$ are determined by electrostatics (Fig.~\ref{sketchnu-2}).
For odd $\nu$, an interlayer phase coherent ($0<\theta\le\pi/2$) state exists for sufficiently small $u$,
which yields\cite{Barlas2} to a layer polarized state ($\theta=0$)  at $\nu=-3$ and $\nu=1$,
and to a sequence of states with partial or full orbital coherence\cite{Cote} at $\nu=-1$ and $\nu=3$.
Notice that $\nu=-3$ and $\nu=3$ are not related by particle-hole symmetry.
This is best understood from Hund's rules\cite{Barlas}: at $\nu=-3$ only one $n=0$ orbital band is occupied, while at $\nu=3$ only one $n=1$ orbital band is empty.
At $\nu=3$ C\^ot\'e \textit{et al.}\cite{Cote} showed that orbitally coherent states dominate the phase diagram,
whose inter-LL excitations are beyond the scope of this study.
The case of $\nu=-3$ is depicted in Fig.~\ref{sketchnu-3}.

Beyond $S_z$ and $P_z$, the magnitudes $P$ or $S$ are quantum numbers at half-filling $\nu=0$.
All excitons include transitions between the non-central levels $|n|\ge2$; the possibility of transitions from, to, or within the central Landau level octet
depends on the ground state, which also resolves the transitions through the exchange self-energy differences; see Figs.~\ref{sketchnu0}, \ref{sketchnu-2} and \ref{sketchnu-3}.

The excitations are grouped by their optical signature.
Due to the small momenta of optical photons, valley flipping modes are optically inactive.
In the $q\to0$ limit $l_z$ becomes a quantum number, and $l_z=\pm1$ applies for single-photon absorption,\cite{infrared} and
$l_z=0,\pm2$ in electronic Raman processes,\cite{raman} with the $l_z=0$ transitions being dominant.
The angular momentum due to the helicity of the photons is transferred\cite{infrared,raman} entirely to the orbital degree of freedom.
Optically inactive modes include Goldstone modes associated with the broken symmetry (outside the scope of our study) and generic dark modes.

\subsection{$\nu=0$}
\label{nulla}

\begin{figure}[htbp]
\begin{center}
\includegraphics[width=\columnwidth, angle=0, keepaspectratio]{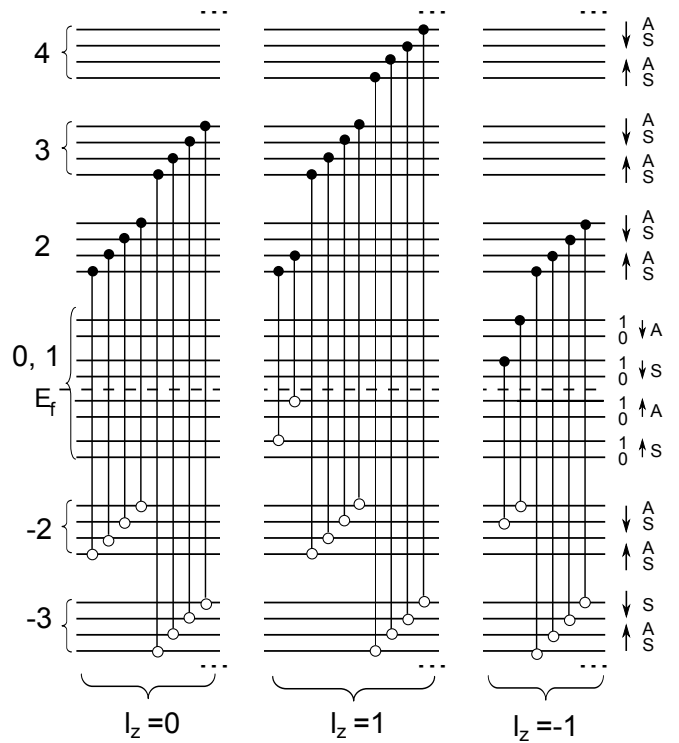}
\end{center}
\caption{\label{sketchnu0}
Magnetoexciton modes at the $\nu=0$ quantum Hall ferromagnetic state in bilayer graphene.
Only the optically relevant spin- and pseudospin-conserving modes are shows.
In the $q\to0$ limit, which is probed by photon absorption or electronic Raman, the transitions in the infinite sequence at fixed $l_z=|n|-|n'|$ may mix;
at $q>0$, the excitations result from the mixing of all $l_z$ sequences.
}
\end{figure}

It is known that two QHF ground states exist, a spin-polarized one and a valley (layer) polarized one\cite{Barlas,TF,Gorbar,Weitz,Kim,Kharitonov,foot3}.
Their respective range of validity is determined by the ratio of the Zeeman energy $\Delta_Z$ to the energy difference between the valleys, which in
turn is related to the potential difference $u$. (In fact, layer and pseudospin can de identified within the central Landau level octet.)
For concreteness, we are discussing the ferromagnetic state.
Here the magnitude of the pseudospin $P$ is a good quantum number of the excitations.
See Fig.~\ref{sketchnu0} for the transitions that span the Hilbert space of the mean-field Hamiltonian.

With the electron ($n$) and hole ($n'$) Landau levels fixed, the $S_z=-1$ transitions consist of a pseudospin triplet\cite{foot}
$\Psi^\dag_{n A\downarrow,n'S\uparrow}$, $\Psi^\dag_{n S\downarrow,n'A\uparrow}$, $\frac{1}{\sqrt2}\left( \Psi^\dag_{n S\downarrow,n'S\uparrow}-\Psi^\dag_{n A\downarrow,n'A\uparrow}\right)$,
and a singlet\cite{foot} $\frac{1}{\sqrt2}\left( \Psi^\dag_{n S\downarrow,S\uparrow}+\Psi^\dag_{n A\downarrow,A\uparrow}\right)$.
This group contains the intralevel transitions among the $n=0,1$ Landau bands; the Goldstone modes associated with the spin rotational symmetry breaking
should be in this subspace.
However, our approach is not appropriate for the description of Goldstone modes even at even filling factors, as we will discuss below.

The $S_z=1$ pseudospin triplet, $\Psi^\dag_{n A\uparrow,n'S\downarrow}$, $\Psi^\dag_{n S\uparrow,n'A\downarrow}$,
$\frac{1}{\sqrt2}\left(\Psi^\dag_{n S\uparrow,n'S\downarrow}-\Psi^\dag_{n A\uparrow,n'A\downarrow}\right)$,
and pseudospin singlet  $\frac{1}{\sqrt2}\left(\Psi^\dag_{n S\uparrow,n'S\downarrow}+\Psi^\dag_{A\uparrow,n'A\downarrow}\right)$, respectively, contains inter-LL transitions only.

\begin{figure}[htbp]
\begin{center}
\includegraphics[width=\columnwidth, angle=0, keepaspectratio]{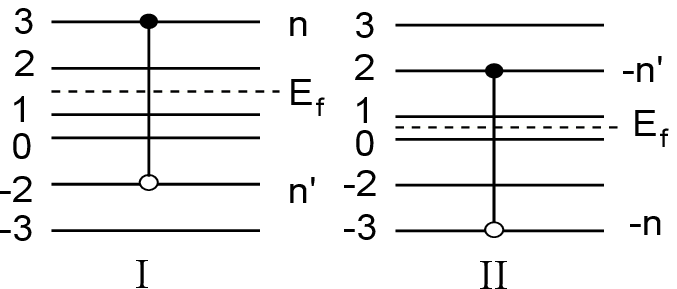}
\end{center}
\caption{\label{phconj}
Imagine two transitions: (I) a particle in level $n'$ is promoted to level $n$ of a valley-spin component (left) of which the $n=1$ Landau band is filled, and (II) $-n\to -n'$
of a component (right) of which the $n=1$ band is empty. 
Particle-hole conjugation of (I) is the promotion of a hole from Landau level $-n'$ to $-n$, i.e., (II); the excitation energies must be the same.
The consequence is Eq.~(\ref{idph1}); the formal proof is straightforward.
}
\end{figure}

The $S_z=0$ sector consists of (i) two triplets,
$\Psi^\dag_{n A\uparrow,n'S\uparrow}$, $\Psi^\dag_{n S\uparrow,n'A\uparrow}$, $\frac{1}{\sqrt2}\left(\Psi^\dag_{n A\uparrow,n'A\uparrow}-\Psi^\dag_{n S\uparrow,n'S\uparrow}\right)$,
and $\Psi^\dag_{n A\downarrow,n'S\downarrow}$, $\Psi^\dag_{n S\downarrow,n'A\downarrow}$, $\frac{1}{\sqrt2}\left(\Psi^\dag_{n S\downarrow,n'S\downarrow}-\Psi^\dag_{n A\downarrow,n'A\downarrow}\right)$,
the RPA terms does not contribute to, and (ii) two singlets,
$\frac{1}{\sqrt2}\left(\Psi^\dag_{n S\uparrow,n'S\uparrow}+\Psi^\dag_{n A\uparrow,n'A\uparrow}\right)$ and
$\frac{1}{\sqrt2}\left(\Psi^\dag_{n S\downarrow,n'S\downarrow}+\Psi^\dag_{n A\downarrow,n'A\downarrow}\right)$,
which are mixed by the RPA term.
Careful inspection reveals, however, that the two pseudospin singlets (ii) always appear in the mean-field Hamiltonian on equal footing, e.g.,
the $\frac{1}{\sqrt2}\left(\Psi^\dag_{2 S\uparrow,1S\uparrow}+\Psi^\dag_{2 A\uparrow,1A\uparrow}\right)$ transition is indistinguishable on the mean-field level from the
$\frac{1}{\sqrt2}\left(\Psi^\dag_{1 S\downarrow,-2,S\downarrow}+\Psi^\dag_{1A\downarrow,-2,A\downarrow}\right)$ transition.
This follows by
\begin{gather}
E_{(2,1)}^{(2,1)}=E_{(1,-2)}^{(1,-2)},\label{ide}\\
R_{(2,1)}^{(2,1)}=R_{(1,-2)}^{(1,-2)}=-R_{(2,1)}^{(1,-2)}=-R_{(1,-2)}^{(2,1)},\label{idr}
\end{gather}
and the following easily provable identity of the exchange self-energy cost:
\begin{multline}
\Delta(n,n')+X_{n'0}+X_{n'1}-X_{n,0}-X_{n,1}= \Delta(-n',-n).\label{idph1}
\end{multline}
(For $n=1$ or $n'=1$ no sign change is necessary.)
Eq.~(\ref{idph1}) simply expresses particle-hole symmetry, i.e., that the exchange self-energy cost of transitions related by particle-hole conjugation
in a fixed component must be identical. See Fig.~\ref{phconj} for the transitions whose comparison yields Eq.~(\ref{idph1}).

The RPA terms are the same in each diagonal and off-diagonal position among equivalent transitions, thus they select the even and the odd
linear combinations in group (ii).
The even combination, $\hat\Psi^{\dag00}_{nn'}(\mathbf q)$ defined in Eq.~(\ref{m0}), gets an RPA enhancement, while the RPA cancels from the alternating sign combinations,
making it energetically equivalent to the $P_z=0$ element of the triplets (i).
Thus, eventually, the $S_z=0,P_z=0$ sector contains a three-fold degenerate curve and a nondegenerate mode.

\begin{figure*}[htbp]
\begin{center}
\includegraphics[width=\textwidth, keepaspectratio]{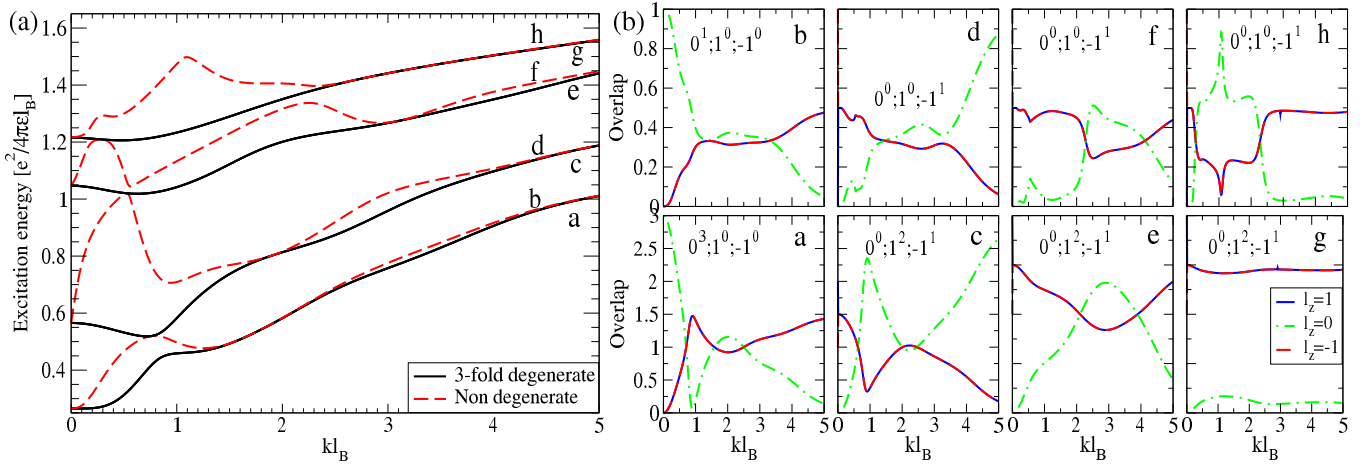}
\end{center}
\caption{\label{spectrumnulla}
(Color online)
(a) Excitation spectra of the quantum Hall ferromagnet at $\nu=0,\pm2$ at $B=10$ T. Only the optically relevant $S_z=P_z=0$ modes are included.
(b) The weight of the shown excitations on the definite $l_z$ subspaces for $\nu=0$ in bottom-up order. For degenerate curves the weight is summed.
The quantum numbers of the $q\to0$ limit are indicated.
}
\end{figure*}

Each of the four multiplets in the $S_z=0$ sector contains a $P_z=0$ mode, which is active in electronic Raman or IR absorption.
Here the mixing of Landau levels results in more widely separated modes.
The optically active excitations are shown in Fig.~\ref{spectrumnulla} with LL mixing taken into account.
Notice that the $l_z=1$ and $l_z=-1$ transitions have an equal weight in all modes, consistent with the particle-hole symmetry at $\nu=0$.

In the $S_z=-1$ sector we find spin waves.
Neglecting Landau level mixing, they give rise to a gapless and a gapped intra-LL modes,\cite{TF} and a sequence of higher inter-LL modes;
each of these is raised by the Zeeman energy and split by the valley energy difference in turn.
The interaction, however, mixes these excitations, thus a clear-cut classification into intra-LL and inter-LL is no longer possible.
Level repulsion unavoidably lowers the formerly gapless modes.
This effect yields apparently negative excitation energies at small wavelength.
Goldstone's theorem, however, ensures that a gapless spin-wave mode is associated with the breaking of the spin rotation symmetry.
Consequently, the seemingly negative energy of the lowermost excitation with a large intra-LL component is an artifact of the combination of Hartree-Fock mean-field theory and LL mixing.
The same anomaly occurs for monolayer graphene,\cite{Iyengar,Lozovik} but it is less apparent when the particle-hole binding energy is plotted.

\subsection{$\nu=\pm2$}

\begin{figure}[htbp]
\begin{center}
\includegraphics[width=\columnwidth, angle=0, keepaspectratio]{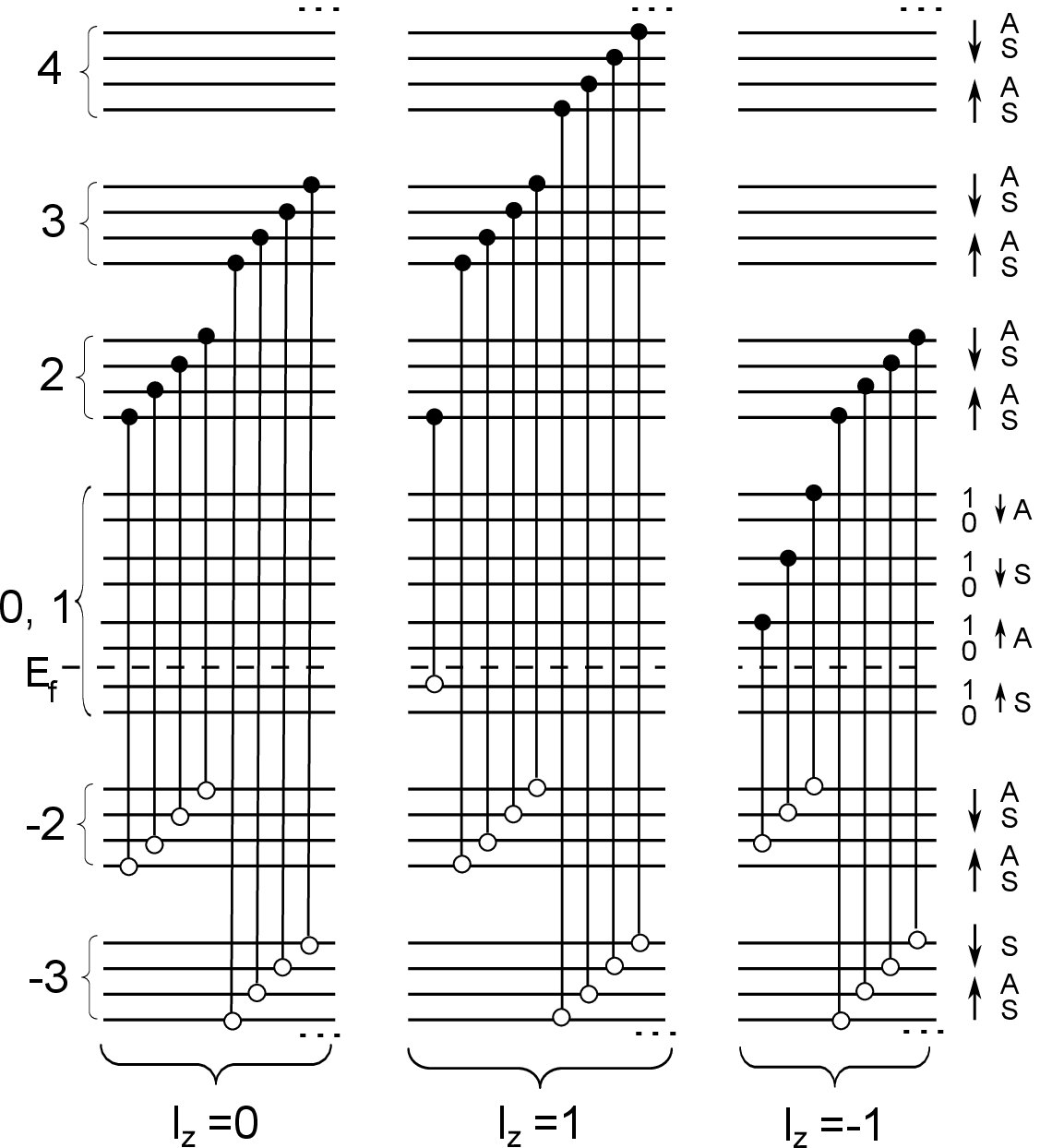}
\end{center}
\caption{\label{sketchnu-2}
The spin- and pseudospin-conserving magnetoexciton modes at the $\nu=-2$ quantum Hall ferromagnetic state in bilayer graphene.
The modes at $\nu=+2$ are obtained by particle-hole conjugation.
}
\end{figure}

The $\nu=\pm2$ state breaks the spin and pseudospin rotational symmetries as the ground state fills the $n=0$ and $n=1$ orbitals of the most favorable spin-pseudospin component, $S\uparrow$.
We restrict the discussion to spin and pseudospin preserving excitations.
See Fig.~\ref{sketchnu-2} for the possible transitions.

\begin{figure}[htbp]
\begin{center}
\includegraphics[width=\columnwidth, keepaspectratio]{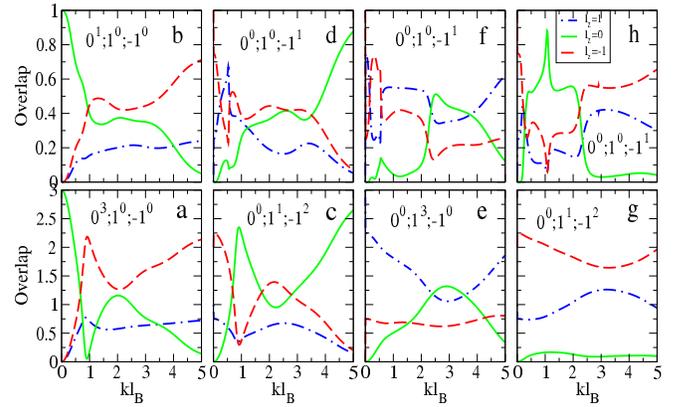}
\end{center}
\caption{\label{spectrumketto}
(Color online)
The projection of the excitations at $\nu=-2$ on the definite $l_z$ subspaces. For degenerate curves the weight is summed.
The spectrum at $\nu=\pm2$ is identical to the one in the left panel of Fig.\ \ref{spectrumnulla}, the letters refer to the same curves.
For $\nu=+2$ the sign of the $+l_z$ and $-l_z$ projections are interchanged w.r.t.\ $\nu=-2$.
}
\end{figure}

It is easy to check that the mean-field Hamiltonian matrix is identical to the one at $\nu=0$.
For the $-n\to n$ transitions ($n\ge2$ integer) this holds because the occupancy of the central Landau level octet is irrelevant as
\begin{equation}
X_{-n,0}+X_{-n,1}-X_{n,0}-X_{n,1}=0.
\end{equation}
The octet of $-(n+1)\to n$ and $-n\to (n+1)$ transitions gives rise to two quartets of equivalent transitions by Eqs.~(\ref{ide}-\ref{idph1}).
While at $\nu=0$ the $S\uparrow$ and $A\uparrow$ transitions of the former group bundle with the $S\downarrow$ and $A\downarrow$ transitions of the second group,
now the $S\uparrow$ transition of the first group bundle with the $A\uparrow$, $S\downarrow$ and $A\downarrow$ transitions of the second group.
The spectrum is still the one in Fig.~\ref{spectrumnulla}(a).
The orbital projection of the modes differs, c.f.\ Fig.~\ref{spectrumketto} for $\nu=-2$.
At $\nu=+2$ the sign of $l_z$ changes in all projections, which determines the helicity of the absorbed and inelastically scattered photons.

\subsection{$\nu=-3$}
\label{minusthree}

\begin{figure}[htbp]
\begin{center}
\includegraphics[width=\columnwidth, angle=0, keepaspectratio]{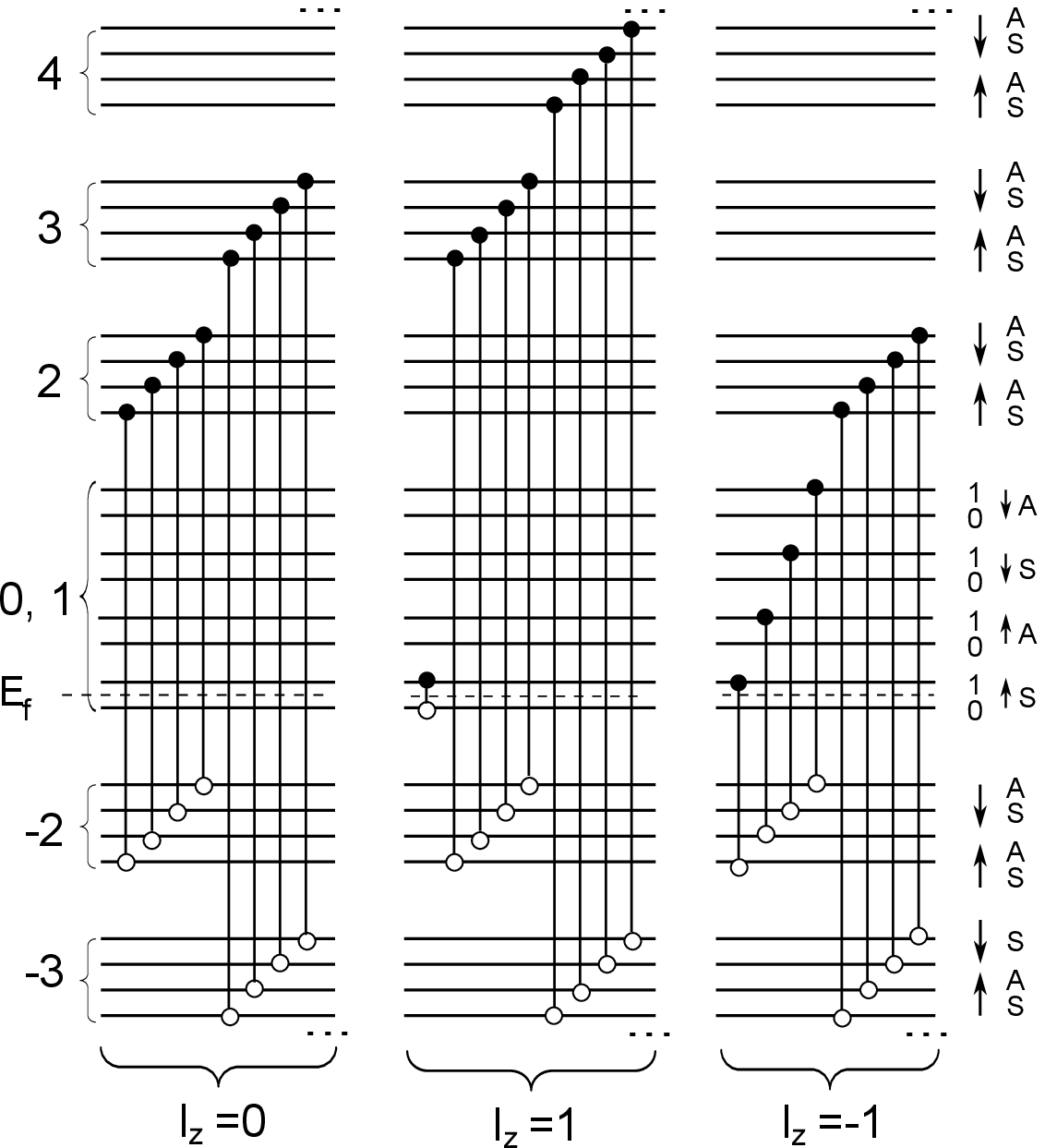}
\end{center}
\caption{\label{sketchnu-3}
The optically spin- and pseudospin-conserving magnetoexciton modes at the $\nu=-3$ quantum Hall ferromagnetic state in bilayer graphene.
Notice that the $S\downarrow$, $A\uparrow$ and $A\downarrow$ transitions occur symmetrically, i.e., the same electron and hole Landau levels are allowed and
the same self-energy cost is picked up from the exchange with the filled levels.
}
\end{figure}

Based on exchange energy considerations within the central Landau level octet, Hund's rule\cite{Barlas} implies that the only occupied band is $(0,S\uparrow)$ at $\nu=-3$.
The states in Eq.~(\ref{defS}) progress from the layer balanced limit $\theta=\pi$ at $u=0$
to the layer polarized state $\theta=0$; this limit is achieved about $u=u_c\approx0.001 e^2/(4\pi\epsilon_0\epsilon_r\ell_B)$, which is only 0.082 meV at $B=20$ T.
For $0<u<u_c$ there is interlayer phase coherence.\cite{Barlas2}
Thus magnetoexcitons exist on both side of $u_c$; the amount electrostatics raises energy of the pseudospin-flipping modes w.r.t.\ the pseudospin conserving modes saturates at $u=u_c$.
Both the spin and the pseudospin symmetries are broken resulting in three Goldstone modes.\cite{Barlas2}
Further, Barlas \textit{et al.}\cite{Barlas2} showed that at finite $u$ there is an instability to a stripe ordered phase with a rather small critical temperature.
Our analysis below applies only below this temperature.
See Fig.~\ref{sketchnu-3} for the transitions that span the Hilbert space of the mean-field Hamiltonian.

We have become aware of Ref.~\onlinecite{Shizuya3}, which derives a modified version of Hund's rule
considering the exchange field due to all filled levels, not just those in the central Landau level octet.
While this makes no difference for even filling factors, at $\nu=-3$ the predicted mean-field ground state fills a band of states which are equal linear
combinations of $n=1$ and $n=0$ Landau orbitals.
The calculation of the excitations of such ground states is delegated to future work.

Because of the degeneracy of $n=0,1$ orbitals, states in central Landau level octet at odd integer fillings involve fluctuations with in-plane electric dipole character.\cite{Barlas}
The consequent collective modes have been studied in detail by Barlas \textit{et al}.\cite{Barlas2} and C\^ot\'e \textit{et al}.\cite{Cote}
As we do not handle such dipolar interactions, we have omitted the predominantly intra-LL lowest curve from the spectra,
and we have checked that the inter-LL excitation modes we keep contain the
$0\to1$ magnetoexcitons with a negligible weight. Reassuringly, we always got a weight less than 0.1\%.

As the $n=0$ orbital is filled with $S\uparrow$ electrons in the mean-field ground state, for fixed electron ($n$) and hole ($n'$) Landau levels
the exchange self-energy cost of the $S\uparrow$ transition is higher than those of the other components.
Also, in the $S\uparrow$ component an intralevel $0\to1$ transition is possible, which mixes with higher $S\uparrow$ transitions;
see Fig.~\ref{sketchnu-3} for the restrictions on the possible transitions at this filling.
The other three components, on the other hand, occur symmetrically in the mean-field Hamiltonian; c.f.\ Fig.~\ref{sketchnu-3}.
One can change basis from the excitons of type $S\downarrow$, $A\uparrow$ and $A\downarrow$ to
\begin{align}
\hat\Psi^{\dag d1}_{nn'}&=\frac{1}{\sqrt 2}\left(\Psi^\dag_{n A\uparrow,n'A\uparrow} - \Psi^\dag_{n A\downarrow,n'A\downarrow}\right),\label{change1}\\
\hat\Psi^{\dag d2}_{nn'}&=\frac{1}{\sqrt 6}\left(\Psi^\dag_{n A\uparrow,n'A\uparrow} + \Psi^\dag_{n A\downarrow,n'A\downarrow}
- 2\Psi^\dag_{n S\downarrow,n'S\downarrow}\right),\\
\hat\Psi^{\dag r}_{nn'}&=\frac{1}{\sqrt 3}\left(\Psi^\dag_{n A\uparrow,n'A\uparrow} + \Psi^\dag_{n A\downarrow,n'A\downarrow}\label{change3}
+ \Psi^\dag_{n S\downarrow,n'S\downarrow}\right).
\end{align}
The RPA contribution cancels from $\hat\Psi^{\dag d1}_{nn'}(\mathbf q)$ and $\hat\Psi^{\dag d2}_{nn'}(\mathbf q)$, which give rise to doubly degenerate excitations.
$\hat\Psi^{\dag r}_{nn'}(\mathbf q)$ has an RPA contribution.
Its mixture with the distinguished $S\uparrow$ excitations produces nondegenerate curves.
In higher energy excitations, on the other hand, the weight of the $0\to1$ transition of the $S\uparrow$ component becomes extremely small,
thus the equivalence of the four components will be approximately restored, yielding threefold quasi-degenerate and nondegenerate curves.

\begin{figure*}[htbp]
\begin{center}
\includegraphics[width=\textwidth, keepaspectratio]{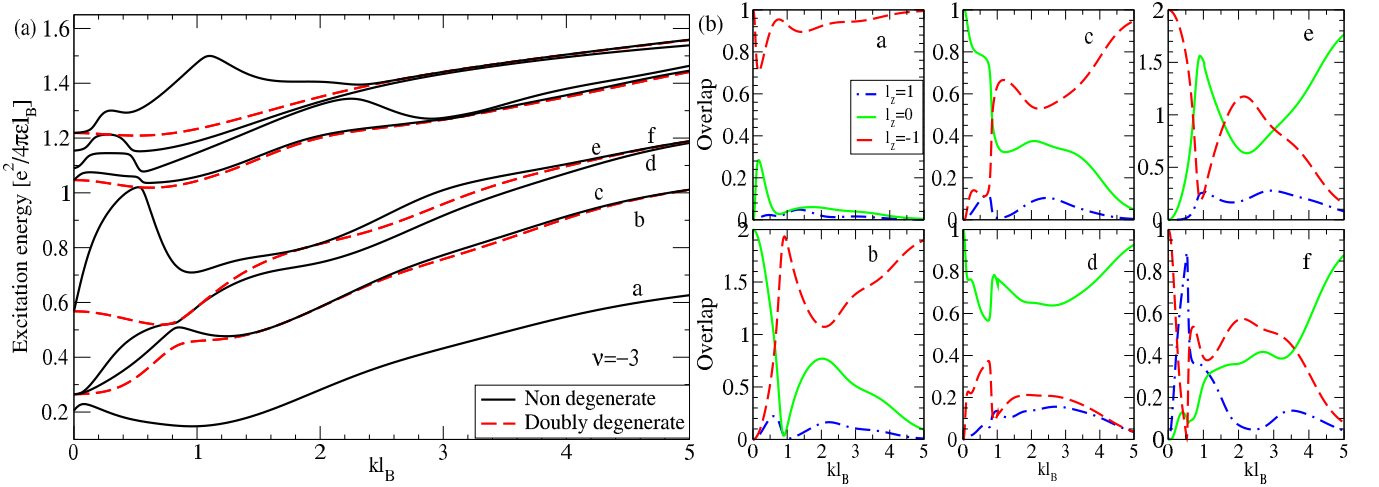}
\end{center}
\caption{\label{spectrumharom}
(Color online)
(a) Excitation spectrum of the quantum Hall ferromagnet in the $\nu=-3$ at $B=10$ T. Only the optically relevant $S_z=P_z=0$ modes are included.
(b) The weight of the shown excitations on the definite $l_z$ subspaces in bottom-up order. For degenerate curves the weight is summed.
}
\end{figure*}

See Fig.~\ref{spectrumharom} for the dispersion of the active excitations.
The small graphs show that the projection to definite $l_z$ subspaces changes abruptly at nonzero wave vector.
Notice that at higher energies $\gtrsim 1 e^2/(4\pi\epsilon_0\epsilon_r\ell_B)$ the double degenerate curves occur in the vicinity of a nondegenerate one,
indicating the approximate restoration of the equivalence of components in this limit.

\subsection{$\nu=1$}
\label{minusone}

Just like at $|\nu|=3$, the state at $\nu=-1$ is not the particle-hole conjugate of the state $\nu=1$.
We do not study $\nu=-1$ because of the relavance of orbitally coherent states.\cite{Barlas3}

At $\nu=1$ the interlayer coherent and layer polarized QHF states both have magnetoexcitons.
The possible transitions in the mean-field Hamiltonian are obtained trivially by raising the Fermi level by four levels in Fig.~\ref{sketchnu-3}.
Now the $S\downarrow$ electrons are distinguished by the possibility of an intralevel $0\to1$ transition, and their higher self-energy.
The other three components occur symmetrically in the mean-field Hamiltonian.
The argument is similar to the case of $\nu=-3$; for the two $1\to2$ transitions of spin-$\uparrow$ electrons and the $-2\to1$ transitions of the spin-$\downarrow$ electrons one uses
$R_{(1,0)}^{(2,1)}=-R_{(1,0)}^{(1,-2)}$ and Eq.~(\ref{idph1}).
When subsequent $l_z=\pm1$ transitions are included, the $-n\to(n+1)$ transition of $A\downarrow$ is equivalent to the $-(n+1)\to n$ transition of $S\uparrow$ and $A\uparrow$,
while the $-n\to(n+1)$ transition of $A\uparrow$  and $S\uparrow$ is equivalent to the $-(n+1)\to n$ transition of $A\downarrow$.
The convenient basis change is similar to Eqs.~(\ref{change1}) to (\ref{change3}), with $S\uparrow$ replaced by $A\downarrow$.

There is only a slight difference between the inter-Landau level excitation spectra at $\nu=-3$ and $\nu=1$: the exchange self-energy cost of the distinguished
transition ($S\uparrow$ at $\nu=-3$ and $S\downarrow$ at $\nu=1$) relative to the three equivalent ones is $X_{n'0}-X_{n0}$ at $\nu=-3$ and $X_{n'1}-X_{n1}$ at $\nu=1$.
As this difference is already small for the lowest transitions and then decreases, we omit the $\nu=1$ spectrum;
its difference from Fig.~\ref{spectrumharom}(a) is comparable to the line width.

\section{Conclusion}
\label{conclusion}

We have calculated the inter-Landau level magnetoexcitons in the integer quantum Hall states as wells as the quantum Hall ferromagnets at filling factor $\nu=-3,\pm2,0,1$ of bilayer graphene.
We have found that the spinorial structure of the orbitals together with the enhanced electron-hole exchange interaction effects in this multicomponent
system gives rise to rather complex dispersions; these are related both to the shape of the Fourier transform of Landau orbitals and to the elimination of crossings by the
mixing of Landau levels, which is significant because the scale of the Coulomb energy is comparable to the cyclotron energies.

The $q\to0$ limit of the excitation can be probed by optical absorption\cite{infrared} and electronic Raman experiments.\cite{raman,ramanexp}
Unlike for the conventional two-dimensional electron gas with a quadratic dispersion,\cite{Kohn} the excitations in a quantizing magnetic field do acquire an interaction shift;
the magnitude of such a shift is one of our experimental predictions.

The wave vector dependence of the magnetoexcitons can be probed in resonant inelastic light scattering experiments,\cite{Pinczuk} where momentum conservation
breaks down mainly because of ineffectively screened charged impurities.
In semiconducting samples this technique has been successfully applied, e.g., to the study of spin-conserving and spin-flipping excitations at integers,\cite{Pinczuk}
of the long wave-length behavior of the low-lying\cite{Pinczuk3} and higher\cite{Pinczuk4} excitations of fractional quantum Hall states,
and of magnetoroton minuma at $\nu=2$ \cite{Pinczuk2} and at fractions.\cite{Pinczuk6}
In graphene, the magnetophonon resonance was observed by this method.\cite{Pinczuk5}

The mixing of transitions that involve different Landau levels is strong in the experimentally accessible range.
This mixing smoothens the dispersion relations via level repulsion, and at finite wavelength causes a strong mixing of the modes that have different angular
momenta in the zero wavevector limit.
In particular, we have found that the anticrossings due to Landau level mixing result in a number of undulations in the magnetoexciton dispersions,
whose van Hove singularities must give a strong signal.\cite{Pinczuk}

On the theory side, we have found that the classification of magnetoexciton modes at finite wave vectors by the ``angular momentum quantum number'' is problematic,
that the screening effects---which we have handled via Landau level mixing---are significant, and that this framework is not quite suitable for studying the Goldstone
modes such as spin waves in the symmetry-breaking quantum Hall ferromagnetic states.

\section*{Acknowledgements}
This research was funded by the Hungarian Scientific Research Funds No. K105149.
C. T. was supported by the SROP-4.2.2/08/1/2008-0011 and SROP-4.2.1.B-10/2/KONV-2010-0002 grants, and the Hungarian Academy of Sciences.

\appendix

\section{The two-body problem in bilayer graphene}

Let us consider the Hamiltonian
\begin{equation}
\label{twobody}
\hat H_2=-\frac{1}{2m}\begin{pmatrix}
0 & (\pi_1^\dag)^2 \\ \pi_1^2 & 0
\end{pmatrix}
-\frac{1}{2m}\begin{pmatrix}
0 & (\pi_2^\dag)^2 \\ \pi_2^2 & 0
\end{pmatrix}
-u(\mathbf r_1-\mathbf r_2),
\end{equation}
where $\pi_{i}=p_{i,x}+ip_{i,y}$, $\mathbf p_1=-i\hbar\nabla_1 -(e/c)\mathbf A(\mathbf r_1)$ belongs to the electron and $\mathbf p_2=-i\hbar\nabla_2 +(e/c)\mathbf A(\mathbf r_2)$ belongs to the hole, and $u(\mathbf r)=e^2/\epsilon r$.
(We have fixed the valley of both the electron and the hole.
A valley-independent interaction is assumed because any deviation from this is small in the ratio of the lattice constant to the magnetic length.)
Introducing center-of-mass and relative coordinates $(X,Y)=\frac{\mathbf r_1+\mathbf r_2}{2}$, $(x,y)=\mathbf r_1-\mathbf r_2$ and momenta $(P_x,P_y)=\mathbf r_1+\mathbf r_2$, $(p_x,p_y)=\frac{\mathbf p_1+\mathbf p_2}{2}$,
and separating the center-of-mass motion by the canonical transformation $\hat U=e^{iXy}$, we obtain
\begin{multline}
\label{transformed}
\hat H_2'=-\frac{1}{m}\begin{pmatrix}
0 & C_-^2 & (C_+^\dag)^2 & 0 \\
(C_-^\dag)^2 & 0 & 0 & (C_+^\dag)^2 \\
C_+^2 & 0 & 0 & C_-^2 \\
0 & C_+^2 & (C_-^\dag)^2 & 0
\end{pmatrix}
-u(\mathbf r-\mathbf{\hat z}\times\mathbf P),
\end{multline}
where the independent harmonic oscillators $C_\pm$ are defined as
\begin{equation}
\label{harmonic}
C_\pm=\frac{(p_x-i\frac{x}{2})\pm i(p_y-i\frac{y}{2})}{2\sqrt2}.
\end{equation}
In complete analogy to the case of the monolayer\cite{Iyengar}, the eigenstates of the kinetic part of $\hat H_2'$ are
\begin{equation}
\Psi_{n_+,n_-}\propto\begin{pmatrix}
S(n_-)\phi_{|n_+|,|n_-|-2}\\
\phi_{|n_+|,|n_-|}\\
S(n_+)S(n_-)\phi_{|n_+|-2,|n_-|-2}\\
S(n_+)\phi_{|n_+|-2,|n_-|}
\end{pmatrix},
\end{equation}
in terms of two-dimensional harmonic oscillator eigenstates,
\begin{equation}
\phi_{nm}(\mathbf r)=\frac{(C_+^\dag)^{(n)}_\xi}{\sqrt{n!}}\frac{(C_-^\dag)^m}{\sqrt{m!}}\frac{1}{\sqrt{2\pi}\ell_B}e^{-r^2/4\ell_B^2}.
\end{equation}
Above $S(n)=0$ is if $n=0$ or $n=1$, else it is sgn($n$).
Thus the kinetic energy operator clearly commutes with the operator
\[
\hat L= C_+^\dag C_+ - C_-^\dag C_-+\begin{pmatrix}
-2 & 0 & 0 & 0\\
0 & 0 & 0 & 0\\
0 & 0 & 0 & 0\\
0 & 0 & 0 & 2
\end{pmatrix},
\]
which returns $l_z$ as an eigenvalue. This operator, however, commutes with the complete $\hat H_2'$ only in the $\mathbf P\to0$ limit.
Thus we can regard the electron-hole bound state as a two-dimensional harmonic oscillator with clockwise ($C_+$) and anti-clockwise ($C_-$) excitations placed in an external confinement potential.
A nonzero center-of-mass motion breaks the rotational symmetry of this confinement and starts to couple the $\Psi_{n_+,n_-}$ states.

Here we closely follow Iyengar \textit{et al.}\cite{Iyengar}
The two-body Hamiltonian in Eq.~(\ref{twobody}) can be written is terms of center-of-mass and relative coordinates and momenta as
\begin{align}
\hat H_2=&-\frac{m_x^+}{2m}\left(\frac{P_x^2}{2}+2p_x^2-\frac{(P_y-x)^2}{2}-2(p_y-X)^2\right)\nonumber\\
&-\frac{m_x^-}{2m}\left(2P_xp_x - 2(p_y-X)(P_y-x)\right)\nonumber\\
&+\frac{m_y^+}{2m}\left(P_x(P_y-x) + 2p_x(p_y-X)\right)\nonumber\\
&+\frac{m_y^-}{2m}\left(P_x(p_y-X)+(p_y-X)P_x+\right.\nonumber\\
&\left.+p_x(Py-x)+(P_y-x)p_x\right)-u(\mathbf r),
\end{align}
where $m_x^\pm=\frac{1}{2}(\sigma_x\otimes1\pm1\otimes\sigma_x)$, and $m_y^\pm=-\frac{1}{2}(\sigma_y\otimes1\pm1\otimes\sigma_y)$.
With the application of the canonical transformation $\hat U=e^{iXy}$, we obtain
\begin{align}
\hat H_2'=&-\frac{m_x^+}{2m}\left(\frac{y^2}{2}+2{p'_x}^2-\frac{x^2}{2}-2{p'_y}^2\right)\nonumber\\
&-\frac{m_x^-}{2m}\left(2yp'_x - 2p'_yx\right)\nonumber\\
&-\frac{m_y^+}{2m}\left(yx - 4p'_xp'_y\right)\nonumber\\
&-\frac{m_y^-}{2m}\left(-yp'_y-p'_yy+p'_xx+xp'_x\right)\nonumber\\
&-u(\mathbf r-\mathbf{\hat z}\times\mathbf P').\label{intermed}
\end{align}
Eq.~(\ref{transformed}) follows by substituting Eq.~(\ref{harmonic}) into Eq.~(\ref{intermed}).
Notice that $\hat H_2'$ in Eq.~(\ref{transformed}) is independent of the transformed center-of-mass coordinates.
Thus $\mathbf P'$ is conserved; in original variables this corresponds to $\mathbf P - e\mathbf r\times\mathbf B$.


\begin{thebibliography}{99}

\bibitem{Novoselov} K. S. Novoselov, E. McCann, S. V. Morozov, V. I. Falko, M. I. Katsnelson, U. Zeitler, D. Jiang, F. Schedin, and A. K. Geim, Nat. Phys. \textbf{2}, 177 (2006).

\bibitem{Bernal} J. D. Bernal, Proc. Roy. Soc. A \textbf{106}, 749 (1924).

\bibitem{review} V. I. Fal'ko, Phil. Trans. R. Soc. A \textbf{366}, 205 (2008).  

\bibitem{Klitzing} K.~von Klitzing, G.~Dorda, and M.~Pepper, Phys.\ Rev.\ Lett. \textbf{45}, 494 (1980).

\bibitem{qhereview} V. I. Fal'ko, Phil. Trans. R. Soc. A \textbf{366}, 205 (2008).

\bibitem{tightbinding} E. McCann and V. I. Fal'ko, Phys.\ Rev.\ Lett. \textbf{96}, 086805 (2006); F. Guinea, A. H. Castro Neto, N. M. R. Peres, Phys.\ Rev.\ B \textbf{73}, 245426 (2006);

\bibitem{Pereira} J. M. Pereira, Jr., F. M. Peeters, and P. Vasilopoulos, Phys.\ Rev.\ B \textbf{76}, 115419 (2007).

\bibitem{magnetoelectric} L. M. Zhang, M. M. Fogler, and D. P. Arovas, Phys. Rev. B \textbf{84}, 075451 (2011).

\bibitem{Kurganova} E. V. Kurganova, A. J. M. Giesbers, R. V. Gorbachev, A. K. Geim, K. S. Novoselov, J. C. Maan, U. Zeitler, Solid State Commun. \textbf{150}, 2209 (2010).

\bibitem{Henriksen1} E. A. Henriksen, Z. Jiang, L.-C. Tung, M. E. Schwartz, M. Takita, Y.-J. Wang, P. Kim, H. L. Stormer, Phys. Rev. Lett. \textbf{100}, 087403 (2008).

\bibitem{ramanexp} C. Faugeras, M. Amado, P. Kossacki, M. Orlita, M. K\"uhne, A. A. L. Nicolet, Yu. I. Latyshev, and M. Potemski, Phys. Rev. Lett. \textbf{107}, 036807 (2011).

\bibitem{Feldman} B. E. Feldman, J. Martin, and A. Yacoby, Nature Physics \textbf{5}, 889 (2009).

\bibitem{Zhao} Y. Zhao, P. Cadden-Zimansky, Z. Jiang, and P. Kim, Phys.\ Rev.\ Lett. \textbf{104}, 066801 (2010).

\bibitem{Weitz} R. T. Weitz, M. T. Allen, B. E. Feldman, J. Martin, and A. Yacoby, Science \textbf{330}, 812 (2010).

\bibitem{Elferen} H. J. van Elferen, A. Veligura, E. V. Kurganova, U. Zeitler, J. C. Maan, N. Tombros, I. J. Vera-Marun, and B. J. van Wees, Phys. Rev. B \textbf{85}, 115408 (2012). 

\bibitem{Martin} J. Martin, B. E. Feldman, R. T. Weitz, M. T. Allen, and A. Yacoby, Phys. Rev. Lett. \textbf{105}, 256806 (2010).

\bibitem{Barlas} Y. Barlas, R. C\^ot\'e, K. Nomura, and A. H. MacDonald, Phys.\ Rev.\ Lett. \textbf{101}, 097601 (2008).

\bibitem{Bao} W. Bao, Z. Zhao, H. Zhang, G. Liu, P. Kratz, L. Jing, J. Velasco, Jr., D. Smirnov, and C. N. Lau, Phys. Rev. Lett. \textbf{105}, 246601 (2010).

\bibitem{McCann} E. McCann, Phys. Rev B \textbf{74}, 161403 (2006).

\bibitem{Ohta} T. Ohta, A. Bostwick, T. Seyller, K. Horn, and E. Rotenberg, Science \textbf{313}, 951 (2006).

\bibitem{Castro} E. V. Castro, K. S. Novoselov, S. V. Morozov, N. M. R. Peres, J. M. B. Lopes dos Santos, J. Nilsson, F. Guinea, A. K. Geim, and A. H. Castro Neto, Phys.\ Rev.\ Lett. \textbf{99}, 216802 (2007).

\bibitem{Min} H. Min, B. Sahu, S. K. Banerjee, and A. H. MacDonald, Phys. Rev. B \textbf{75}, 155115 (2007).

\bibitem{Oostinga} J. B. Oostinga, H. B. Heersche, X. Liu, A. F. Morpurgo, and  L. M. K. Vandersypen, Nat. Mat. \textbf{7}, 151 (2007) .

\bibitem{Kuzmenko} A. B. Kuzmenko, E. van Heumen, D. van der Marel, P. Lerch, P. Blake, K. S. Novoselov, and A. K. Geim, Phys. Rev. B \textbf{79}, 115441 (2009).

\bibitem{Zhang} Y. Zhang, T.-T. Tang, C. Girit, Z. Hao, M. C. Martin, A. Zettl, M. F. Crommie, Y. R. Shen, and F. Wang, Nature \textbf{459}, 820 (2009).

\bibitem{Mak} K. F. Mak, C. H. Lui, J. Shan, T. F. Heinz, Phys. Rev. Lett. \textbf{102}, 256405 (2009).

\bibitem{Mucha} M. Mucha-Kruczy\'nski, E. McCann, V.I. Fal'ko, Solid State Commun. \textbf{149}, 1111 (2009).

\bibitem{Wang} D. Wang and G. Jin, Europhys. Lett. \textbf{92}, 57008 (2010).

\bibitem{dualgate1} J. B. Oostinga, H. B. Heersche, X. Liu, A. F. Morpurgo, and  L. M. K. Vandersypen, Nat. Mat. \textbf{7}, 151 (2007).

\bibitem{dualgate2} Y. Zhang, T.-T. Tang, C. Girit, Z. Hao, M. C. Martin, A. Zettl, M. F. Crommie, Y. R. Shen, and  F. Wang, Nature \textbf{459}, 820 (2009).

\bibitem{dualgate3} J. Yan and M. S. Fuhrer, Nano Lett. \textbf{10}, 4521 (2010).

\bibitem{Kim} S. Kim and E. Tutuc, arXiv:0909.2288 (2009); S. Kim, K. Lee, and E. Tutuc, Phys. Rev. Lett. \textbf{107}, 016803 (2011).

\bibitem{Gorbar}  E. V. Gorbar, V. P. Gusynin, and V. A. Miransky, JETP Lett. \textbf{91}, 314 (2010); Phys.\ Rev.\ B \textbf{81}, 155451 (2010);
E. V. Gorbar, V. P. Gusynin, V. A. Miransky, and I. A. Shovkovy, Phys. Rev. B \textbf{85}, 235460 (2012).

\bibitem{Nand} R. Nandkishore and L. Levitov, Phys. Rev. Lett. \textbf{104}, 156803 (2010); A relevant discussion is only avalable in a previous version of this paper, arXiv:0907.5395v1 (2009),

\bibitem{TF} C. T\H oke and V. I. Fal'ko, Phys. Rev. B \textbf{83}, 115455 (2011).

\bibitem{Kharitonov} M. Kharitonov, Phys. Rev. Lett. \textbf{109}, 046803 (2012).

\bibitem{Kallin} C. Kallin and B. I. Halperin, Phys.\ Rev.\ B \textbf{30}, 5655 (1984).

\bibitem{MacD} A. H. MacDonald, J. Phys. C: Solid State Phys. \textbf{18}, 1003 (1985).

\bibitem{Lerner} I. V. Lerner and Yu. E. Lozovik, Zh. Eksp. Teor. Fiz. \textbf{78}, 1167 (1980) [Sov. Phys. JETP \textbf{51}, 588 (1981)].

\bibitem{Bychkov} Yu. A. Bychkov, S. V. Iordanskii, and G. M. Eliashberg, Pis'ma Zh. Eksp. Teor. Fiz. \textbf{33}, 152 (1981) [Sov. Phys. JETP Lett. \textbf{33}, 143 (1981)];
Yu. A. Bychkov, E. I. Rashba, Zh. Eksp. Teor. Fiz. \textbf{85}, 1826 (1980) [Sov. Phys. JETP \textbf{58}, 1062 (1983)].

\bibitem{infrared} D. S. L. Abergel and V. I. Fal'ko, Phys. Rev. B \textbf{75}, 155430 (2007).

\bibitem{Pinczuk} A. Pinczuk, B. S. Dennis, L. N. Pfeiffer, K. W. West, Semicond. Sci. Technol. \textbf{9}, 1865 (1994).
\bibitem{Pinczuk3} A. Pinczuk, B. S. Dennis, L. N. Pfeiffer, and K. West, Phys. Rev. Lett. \textbf{70}, 3983 (1993).
\bibitem{Pinczuk4} T. D. Rhone, D. Majumder, B. S. Dennis, C. Hirjibehedin, I. Dujovne, J. G. Groshaus, Y. Gallais, J. K. Jain, S. S. Mandal, Aron Pinczuk, L. Pfeiffer, and K. West,
Phys. Rev. Lett. \textbf{106}, 096803 (2011). 
\bibitem{Pinczuk2} A. Pinczuk, S. Schmitt-Rink, G. Danan, J. P. Valladares, L. N. Pfeiffer, and K. W. West, Phys. Rev. Lett. \textbf{63}, 1633 (1989);
M. A. Eriksson, A. Pinczuk, B. S. Dennis, S. H. Simon, L. N. Pfeiffer, and K. W. West, Phys. Rev. Lett. \textbf{82}, 2163 (1999).
\bibitem{Pinczuk6} M. Kang, A. Pinczuk, B. S. Dennis, L. N. Pfeiffer, and K. W. West, Phys. Rev. Lett. \textbf{86}, 2637 (2001). 
\bibitem{Pinczuk5} J. Yan, S. Goler, T. D. Rhone, M. Han, R. He, P. Kim, V. Pellegrini, and A. Pinczuk, Phys. Rev. Lett. \textbf{105}, 227401 (2010).

\bibitem{raman} M. Mucha-Kruczy\'nski, O. Kashuba, and V. I. Fal'ko, Phys. Rev. B \textbf{82}, 045405 (2010).

\bibitem{Yang} K. Yang, S. Das Sarma, A. H. MacDonald, Phys.\ Rev.\ B \textbf{74}, 075423 (2006).

\bibitem{Iyengar} A. Iyengar, J. Wang, H. A. Fertig, and L. Brey, Phys.\ Rev.\ B \textbf{75}, 125430 (2007).

\bibitem{Bychkov2} Yu. A. Bychkov and G. Martinez, Phys. Rev. B \textbf{77}, 125417 (2008).

\bibitem{Roldan} R. Rold\'an, J.-N. Fuchs, and M. O. Goerbig, Phys. Rev. B \textbf{82}, 205418 (2010). 

\bibitem{Lozovik} Yu. E. Lozovik and A. A. Sokolik, Nanoscale Res. Lett. \textbf{7}, 134 (2012).

\bibitem{Kohn} W. Kohn, Phys. Rev. \textbf{123}, 1242 (1961).

\bibitem{Shizuya2} K. Shizuya, Phys. Rev. B \textbf{81}, 075407 (2010).

\bibitem{Jiang} Z. Jiang, E. A. Henriksen, L. C. Tung, Y.-J. Wang, M. E. Schwartz, M. Y. Han, P. Kim, and H. L. Stormer, Phys. Rev. Lett. \textbf{98}, 197403 (2007).

\bibitem{Henriksen2} E. A. Henriksen, P. Cadden-Zimansky, Z. Jiang, Z. Q. Li, L.-C. Tung, M. E. Schwartz, M. Takita, Y.-J. Wang, P. Kim, H. L. Stormer, Phys.\ Rev.\ Lett. \textbf{104}, 067404 (2010).

\bibitem{Deacon} R. S. Deacon, K.-C. Chuang, R. J. Nicholas, K. S. Novoselov, and A. K. Geim, Phys. Rev. B \textbf{76}, 081406(R) (2007).

\bibitem{Zhu} K. Zou, X. Hong, J. Zhu, Phys. Rev. B \textbf{84}, 085408 (2011).

\bibitem{Barlas2} Y. Barlas, R. C\^ot\'e, J. Lambert, A. H. MacDonald, Phys. Rev. Lett. \textbf{104} 096802 (2010).

\bibitem{Cote} R. C\^ot\'e, J. Lambert, Y. Barlas, and A. H. MacDonald, Phys. Rev. B \textbf{82}, 035445 (2010).

\bibitem{Shizuya1} K. Shizuya, Phys. Rev. B \textbf{79}, 165402 (2009); Physica E \textbf{42}, 736 (2010); Phys. Rev. B \textbf{84}, 075409 (2011).

\bibitem{SWM} P. R. Wallace, Phys. Rev. \textbf{71}, 622 (1947); J. W. McClure, Phys. Rev. \textbf{108}, 612 (1957);
J. C. Slonczewski and P. R. Weiss, Phys. Rev. \textbf{109}, 272 (1958).

\bibitem{Zhang2} L. M. Zhang, Z. Q. Li, D. N. Basov, M. M. Fogler, Z. Hao and M. C. Martin, Phys. Rev. B \textbf{78}, 235408 (2008).

\bibitem{Li} Z. Q. Li, E. A. Henriksen, Z. Jiang, Z. Hao, M. C. Martin, P. Kim, H. L. Stormer, and D. N. Basov, Phys. Rev. Lett. \textbf{102}, 037403 (2009).

\bibitem{foot2} Sometimes the intra-Landau level excitons are called spin waves or pseudospin waves, depending on the quantum numbers that
distinguish the filled and the empty levels.
The inter-LL excitons that conserve all quantum numbers are called magnetoplasmons,\cite{Bychkov2} and those that do not are dubbed spin-flip,
valley-flip, or pseudospin-flip excitations.\cite{Roldan}
We do not use these terms; the class of magnetoexcitons include all of these varieties.

\bibitem{skyrmions} S. L. Sondhi, A. Karlhede, S. A. Kivelson, and E. H. Rezayi, Phys.\ Rev.\ B \textbf{47}, 16419 (1993);
H. A. Fertig, L. Brey, R. C\^ot\'e, A. H. MacDonald, A. Karlhede, and S. L. Sondhi, Phys.\ Rev.\ B \textbf{55}, 10671 (1997).

\bibitem{Shizuya3} K. Shizuya, Phys. Rev. B \textbf{86}, 045431 (2012).

\bibitem{Misumi} T. Misumi and K. Shizuya, Phys. Rev. B \textbf{77}, 195423 (2008).

\bibitem{foot} One can easily check that the equal sign linear combination of the excitons created by the operators $(S\downarrow,S\uparrow)$ and $(A\downarrow,A\uparrow)$
is a pseudospin singlet, whereas the opposite sign linear combination is a member of a pseudospin triplet:
Compare $P_-\hat\Psi^\dag_{n\uparrow S,n'\uparrow S}|\text{gs}\rangle$ with $\left(\Psi^\dag_{n\uparrow S,n'\downarrow S}\pm \Psi^\dag_{n\uparrow A,n'\downarrow A}\right)|\text{gs}\rangle$,
where $P_-$ is the total pseudospin lowering operator and $|\text{gs}\rangle$ is a ground state that is invariant for pseudospin rotation.

\bibitem{Moska} S. A. Moskalenko, M. A. Liberman, P. I. Khadzhi, E. V. Dumanov, I. V. Podlesny, and V. Botan, Solid State Commun. \textbf{140}, 236 (1996); Physica E \textbf{39}, 137 (2007).

\bibitem{foot1}
At $q=0$ we have attempted an extrapolation of the excitation energies as a function of the cutoff $M$.
The energies show a decreasing tendency, but fitting a power function\cite{Lozovik} has proved to be impossible.
We have chosen an ad-hoc cutoff at $M=7$, with the understanding that small quantitative deviations are possible, especially at low energies.
We have checked that using $M=15$ does not fundamentally change the spectra.

\bibitem{future} J. S\'ari and C. T\H oke, in preparation.

\bibitem{Bisti} V. E. Bisti and N. N. Kirova, Phys. Rev. B \textbf{84}, 155434 (2011).

\bibitem{foot3} At certain values of the in-plane pseudospin anisotropy, which may arise due to electron-electron and electron-phonon interactions, two more phases are possible:
a canted antiferromagnet and a partially layer polarized state, c.f.\ Ref.~\onlinecite{Kharitonov}. These recently proposed states are beyond the scope of our study.

\bibitem{Barlas3} Y. Barlas, W.-Ch. Lee, K. Nomura, and A. H. MacDonald, Int. J. Mod. Phys. B \textbf{23}, 2634 (2009).

\end{thebibliography}
\end{document}